\documentclass[a4paper,11pt]{article}

\usepackage[margin=1.14in]{geometry} 
\usepackage{tabularx}
\usepackage{multirow}
\usepackage{hyperref}
\usepackage{booktabs}
\usepackage{arydshln}
\usepackage{amsmath}
\usepackage{amsthm}
\usepackage{amssymb}
\usepackage{graphicx}
\usepackage{pdfpages}
\usepackage{float}
\usepackage{bm}
\makeatletter
\makeatletter \renewcommand{\@dotsep}{10000} \makeatother
\usepackage{wrapfig}
\usepackage[utf8]{inputenc}
\usepackage{lmodern}
\usepackage{hyperref}
\usepackage{booktabs}  % in your preamble
\usepackage{authblk}

\definecolor{darkred}{rgb}{0.6, 0, 0}
\hypersetup{
	unicode,
	colorlinks,
	breaklinks,
	urlcolor=darkred, 
        linkcolor=black, 
        citecolor=darkred,
	pdfauthor={Author One, Author Two, Author Three},
	pdfproducer={LaTeX},
	pdfcreator={pdflatex}
}
\usepackage[T1]{fontenc} % if needed
\usepackage{amsmath}

\newcommand{\be}{\begin{eqnarray}}
\newcommand{\ee}{\end{eqnarray}}
\def\be{\begin{equation}}
\def\ee{\end{equation}}
\def\bea{\begin{eqnarray}}
\def\eea{\end{eqnarray}}

\newcommand{\gsim}{\;\raisebox{-0.9ex}{$\textstyle\stackrel{\textstyle >}{\sim}$}\;}
\newcommand{\lsim}{\;\raisebox{-0.9ex}{$\textstyle\stackrel{\textstyle<}{\sim}$}\;}
\def\lsim{\raise0.3ex\hbox{$\;<$\kern-0.75em\raise-1.1ex\hbox{$\sim\;$}}}
\def\gsim{\raise0.3ex\hbox{$\;>$\kern-0.75em\raise-1.1ex\hbox{$\sim\;$}}}

\usepackage{graphics}

\usepackage{epsfig}
\usepackage{slashed}
\usepackage[utf8]{inputenc}
\usepackage{changepage}
\usepackage{multirow}
\usepackage{dcolumn}

\usepackage[sort&compress,numbers,square]{natbib}
\theoremstyle{plain}

\theoremstyle{definition}

\usepackage{graphicx, color}

\title{\bf Explaining  Data Anomalies over the NMSSM Parameter Space with Deep Learning Techniques} 

\vspace{8mm}
\author[1]{\Large A.~Hammad}
\author[2]{\Large Raymundo~Ramos}
\author[3]{\\\Large Amit~Chakraborty}
\author[4]{\Large Pyungwon~Ko}
\author[5,6]{\Large Stefano~Moretti}
\affil[1]{\small Theory Center, IPNS, KEK,  1-1 Oho, Tsukuba, Ibaraki 305-0801, Japan}
\affil[2]{\small Quantum Universe Center, KIAS, Seoul 02455, Korea}
\affil[3]{\small Department of Physics, SRM University-AP, Amaravati, Mangalagiri 522240, India}
\affil[4]{\small School of Physics, KIAS, Seoul 02455, Korea}
\affil[5]{\small School of Physics \& Astronomy, University of Southampton, Southampton SO17 1BJ, UK}
\affil[6]{\small Department of Physics \& Astronomy, Uppsala University, Box 516, 75120 Uppsala, Sweden}

\date{}

\begin{document}
	\maketitle
	\vspace{4mm}
	\begin{abstract}
%    By using an efficient numerical scanner package exploiting Deep Learning (DL) techniques, we show how the Next-to-Minimal Supersymmetric Standard Model (NMSSM) can be used as a testbed to capture many of the prevalent data anomalies currently existing in both  low and high energy particle physics, including the so-called 95 and 650 GeV excesses emerging from Higgs searches, the Electro-Weakino excess stemming from searches for Supersymmetry (SUSY), the latest $(g-2)_\mu$ measurements as well as discrepancies from SM predictions existing in potential mono-$Z$ and -$H$ signatures of Dark Matter (DM). We prove that parameter regions exist in this SUSY model where all such dataset features can be accommodated at the $2\sigma$ level, while being complaint with all up-to-date theoretical and experimental constraints. We finally present a few BPs (BPs), wherein all of the above is realised, amenable to further phenomenological investigation. 

%%% AH: I would prefer giving the credit to how the NMSSM can explain the current anomalies not to the scanning method. Here is a slight modified version.

Motivated by recent results from particle physics analyses, we investigate the Next-to-Minimal Supersymmetric Standard Model (NMSSM) as a framework capable of accommodating a range of current data anomalies across low- and high-energy experiments. These include the so-called 95~GeV and 650~GeV excesses from Higgs studies, the Electro-Weakino excess from Supersymmetry searches, the latest $(g-2)_\mu$ measurements as well as potential deviations from Standard Model (SM) predictions that would appear as a consequence in mono-$H$ (where $H=h_{\rm SM}$) and -$Z$  signatures of Dark Matter. Our analysis demonstrates that viable NMSSM parameter regions exist where all these features can be accommodated  at the $2\sigma$ level while remaining consistent with the most up-to-date theoretical and experimental constraints. To identify such regions, we employ an efficient numerical scanning strategy assisted by Deep Learning techniques. We further present several Benchmark Points that realize these scenarios, offering promising directions for future phenomenological studies.

 \normalsize{}

 \end{abstract}

\newpage
\noindent\rule{\textwidth}{1pt}
\tableofcontents
\noindent\rule{\textwidth}{0.2pt}
\maketitle \flushbottom
\vspace{4mm}
%%================================+%%
\section{Introduction} 
%%================================+%%

Leaving aside much experimental evidence pointing to the fact that the Standard Model (SM)  cannot be a {\sl complete} description of Nature (e.g., the existence of 
neutrino masses, Dark Matter (DM) and the matter-antimatter asymmetry of the
Universe),  
from a mere theoretical point of view, it is also clear that the 
SM cannot be a realization of the {\sl  ultimate} theory that is {stable} to high
energies either ({e.g.}, up to the Planck scale, $M_{\rm Planck} \sim 1.2 \times
10^{19}$~GeV, where gravity is expected to become strong and possibly unify with the
other interactions). In fact, there are theoretical inconsistencies leading to instabilities 
of the Higgs boson self-interactions (or, equivalently, of the associate mass) that systematically grow with the energy at which the SM is being probed,
ultimately leading to either its breakdown or a solution perceived as highly {ad-hoc}, hence 
unnatural. This conceptual problem is known as the `hierarchy problem'. 
It is associated with the absence
of a symmetry protecting the Higgs mass, which is necessarily placed at the Electro-Weak (EW) scale (as now confirmed by experiment), 
when the interactions of the Higgs field are instead generated where
the natural cutoff scale is or indeed above it,
i.e., the energy at which a Grand Unification Theory (GUT) 
(also including gravity) will inevitably 
have to be formulated, $M_{\rm GUT}
\simeq 3 \times 10^{16}$~GeV. 

Thus, neither experimentally, nor theoretically, 
the SM can be considered  a {\sl fundamental} theory of Nature and we
need to search for new physics constructions beyond it, which can nonetheless
reproduce the SM in the phenomenological situations where it has been probed to
a phenomenal degree of accuracy. One of
the best candidates for Beyond the SM (BSM) physics is Supersymmetry (SUSY), which is nothing but the most general symmetry of the Lorentz-Einstein theory which relates bosons and fermions, hence implying  a new kind of unification between particles of different spin.

Despite the
absence of experimental verifications of such a theory, relevant theoretical
arguments can
be given in favour of it, i.e., unlike the SM: (i) the SUSY Higgs potential is generated dynamically; (ii) SUSY predicts the existence of a Higgs boson at the EW scale, like the one discovered in 2012; (iii) SUSY facilitates the convergence of the three gauge coupling `constants' at high scale; (iv) the Lightest Supersymmetric Particle (LSP)  can be a natural
candidate for DM, typically the lightest so-called neutralino. 

Thus SUSY is an elegant theory and, like the SM, needs not be minimal, neither in its gauge and Higgs content nor in its matter spectrum~\cite{Moretti:2019ulc}. Indeed, one of the most attractive non-minimal SUSY models is the
Next-to-Minimal Supersymmetric SM (NMSSM) (see Refs.~\cite{Ellwanger:2009dp,Maniatis:2009re} for reviews),
which extends the time-honoured MSSM by the introduction of just one singlet Superfield,
$\hat S$, in addition to the two doublets of the MSSM, $\hat H_u$ and $\hat H_d$. When the scalar component of $\hat S$
acquires a TeV scale Vacuum Expectation Value (VEV), 
the Superpotential term $\hat S \hat H_u \hat H_d$
generates an `effective' so-called $\mu$ term, i.e., $\lambda <S>\hat H_u \hat H_d$, in the form of an  interaction for the Higgs doublet fields. 
Such a term is essential for acceptable 
phenomenology and is put in by hand in the MSSM (in turn giving rise to the $\mu$ problem of the latter). Furthermore, the presence of the additional Higgs state and its VEV also means that the tree-level mass of the aforementioned discovered Higgs boson is (in the NMSSM) much larger than in the MSSM, so that it is more natural to achieve a value for it in the vicinity of 125~GeV, without fine-tuning the loop contributions to it due (primarily) to the {heavy} top squarks (or stops), the notorious little (or small)  hierarchy problem of the MSSM. Thus, the phenomenological implications
of the NMSSM ought to be considered very seriously. 

One aspect of all this is the fact that the Higgs sector
of the MSSM is extended, so that in the NMSSM there are three CP-even Higgs bosons
($H_{1,2,3}$, $m_{H_1}<m_{H_2}<m_{H_3}$), two CP-odd Higgs bosons ($A_{1,2}$,
$m_{A_1}<m_{A_2}$) 
and a charged Higgs pair ($H^\pm$)\footnote{We assume that
CP is not violated in the Higgs sector.}. Another is that the SUSY counterpart of the additional singlet is an additional neutralino, altering the DM phenomenology of the model with respect to the MSSM. Then, there is the vast number and the sheer variety of sparticles entering at one-loop virtually any physics process. An important question
is then whether such a model could be behind a long list of data anomalies currently present in various experimental datasets, especially considering that they have emerged in the context of searches for additional Higgs states (beyond the 125~GeV one), sparticle searches (in particular, non-colored ones) as well as precision measurements of the anomalous magnetic moment of the muon ($(g-2)_\mu$).
It is entirely possible to make this assessment with a high level of precision,
given the sophisticated tools available to make predictions in the context of the NMSSM, ranging from programs computing production  cross-sections and decay rates~\cite{Ellwanger:2004xm,Ellwanger:2005dv,Ellwanger:2006rn}, to protocols interfacing generated NMSSM events to parton shower, hadronization, heavy flavor decay as well as detector software~\cite{Allanach:2008qq}. Finally, structured collaborations across theory and experiment exist to define NMSSM benchmark points (BPs) which can be tested at present and future facilities, starting from Ref.~\cite{Djouadi:2008uw} and ending with the most recent efforts within, e.g.,  the NMSSM Subgroup of the Large Hadron Collider Higgs Cross-Section Working Group 3 (LHC-HXSWG3)\footnote{See {\tt https://twiki.cern.ch/twiki/bin/view/LHCPhysics/LHCHWGNMSSM.}}.

Specifically, the data anomalies for which we seek an NMSSM explanation are as follows. 
CMS has observed mild excesses in the diphoton channel near  95.4~GeV with about $2\sigma$ significance, while ATLAS reported a smaller $1.7\sigma$ excess at 95~GeV~\cite{CMS:2015ocq,CMS:2018cyk,CMS:2024yhz,ATLAS:2024bjr}. Another resonant excess decaying into SM like Higgs and  a scalar of mass $95$ GeV has been observed in $\gamma\gamma \bar{b}b$~\cite{CMS:2023boe}, $\tau^+\tau^- \bar{b}b$~\cite{CMS:2021yci} and $\gamma\gamma \tau^+\tau^-$~\cite{CMS:2025tqi} final states.
The phenomenology of these scalars has received some attention recently~\cite{Choi:2019yrv,Cao:2016uwt,Cao:2019ofo,Biekotter:2021qbc,Li:2022etb,Banik:2023ecr,Ellwanger:2023zjc,Cao:2023gkc,Cao:2024axg,Ellwanger:2024txc,Ellwanger:2024vvs,Lian:2024smg,LeYaouanc:2025mpk}.
Another notable anomaly arises from a mild excess of events reported by both ATLAS~\cite{ATLAS:2019lng,ATLAS:2021moa} and CMS~\cite{CMS:2021edw,CMS:2021far} over the SM expectations: these events feature final states with multiple leptons and large Missing Transverse Energy (MET or $\slashed{E}_T$). In particular, they are consistent with the so-called Electro-Weakino (EWino) excess, i.e., chargino--neutralino pair production, where the subsequent decays of the chargino and neutralino proceed via off-shell \(W^\pm\) and \(Z\) bosons, resulting in soft leptons. Lastly, we take into account the recent measurement highlighting the discrepancy between the experimentally measured muon anomalous magnetic moment~\cite{Muong-2:2006rrc}  and the SM prediction. We aim at explaining all such observations in the context of the NMSSM, in turn predicting anomalies in DM signals (specifically, mono-$H$ and -$Z$ ones, with $H$ and $Z$ being the Higgs and neutral weak boson of the SM, respectively), that will therefore act as collateral predictions of our analysis.

%{\textcolor{violet}{
Moreover, our numerical study starts with a scan of the parameter space of the NMSSM
that employs the recently released package {\tt DLScanner}~\cite{Hammad:2024tzz},
a tool designed to perform scans of parameter spaces in high energy physics models.
It uses deep learning (DL) techniques to identify regions in the parameter space
where a target (conditions, constraints, extrema or other) is likely reached.
It is an appropriate tool for this case as it can be tuned to find target regions
that are complicated or small.
One characteristic of {\tt DLScanner} that helps with small targets
is the use of \texttt{VEGAS}~\cite{Lepage:2020tgj} maps that
can be trained to generate random distributions that are more weighted towards the target.
Considering that {\tt DLScanner} is relatively new,
this study serves as both a test and a demonstration of its capabilities.

%In doing a  scan the NMSSM parameter space in the search of those region of it explaining all the aforementioned anomalies {\sl simultaneously}, we will use the program {\tt DLScanner} \cite{Hammad:2024tzz},  which was deployed by some of us as an efficient  scanner package enhanced by Deep Learning (DL) techniques. 
%{\textcolor{red}{SM: Ahmed/Raymundo to describe the tool a bit more.}} {\textcolor{violet}{[RR: Description extended, check if this is better]}}

Our paper is organized as follows. In Sect.~\ref{sec:NMSSM} we introduce the NMSSM. In the two following ones, we list the datasets that will be used to test such a theoretical model, covering  signatures emerging in both the visible (Higgs, gauge and matter) and invisible (SM) sectors of the NMSSM. Then, we will present and discuss our results in Sect.~\ref{sec:results}. Finally, we will conclude.

%%================================+%%
\section{The NMSSM} 
\label{sec:NMSSM}
%%================================+%%
The NMSSM~\cite{Fayet:1974pd,Dine:1981rt,Nilles:1982dy,Frere:1983ag,Derendinger:1983bz} is an extension of the Minimal Supersymmetric Standard Model (MSSM) that introduces an additional gauge singlet Superfield. (See also Refs.~\cite{Dedes:2000jp,Panagiotakopoulos:2000wp} for alternative SUSY  realizations with the same particle content.) This addition provides a solution to the $\mu$-problem of the MSSM and can naturally explain the mass of the 125 GeV Higgs boson without the need for excessively heavy stop quarks. Primarily, the NMSSM offers a broader range of possibilities for DM candidates and collider signals due to its extended neutralino and Higgs sectors, respectively, as discussed below. 

The NMSSM includes a single-gauge chiral Superfield $\hat{S} $ in addition to the usual two MSSM-type Higgs doublets $\hat{H_u}$ and $\hat{H_d}$, giving rise to additional Higgs and neutralino states compared to their MSSM counterparts. The $Z_3$ invariant NMSSM Superpotential takes the following form: 
\begin{equation} 
W_{\rm NMSSM} = W_{\rm MSSM} + \lambda \hat{H_u} \hat{H_d} \hat{S} + \frac{1}{3} \kappa \hat{S}^3, 
\end{equation}
where $W_{\rm MSSM}$ is the MSSM Superpotential with the $\mu$ parameter set to zero. Once the scalar component of the Superfield $\hat{S}$ develops a VEV, $s\equiv\langle S\rangle$, the second term in the above equation provides the effective $\mu$ term: $\mu_{\rm eff} = \lambda s$. Also notice that, without the 
{$\kappa$} term, the Superpotential above would have a $U(1)'$ symmetry, so-called Peccei--Quinn symmetry~\cite{Peccei:1977hh,Peccei:1977ur}. The role of the 
{$\kappa$} term is to break this $U(1)'$ symmetry. Further, this term is introduced such that 
{$\kappa$} is dimensionless. However, there remains a discrete 
$Z_3$
symmetry, which can finally be broken spontaneously without cosmological consequences (in terms of domain walls)~\cite{Zeldovich:1974uw,Panagiotakopoulos:1998yw}.

The soft SUSY breaking part of the Lagrangian is given by 
\begin{equation}
V_{\rm soft} = m^2_{H_u}|H_u|^2 + m^2_{H_d}|H_d|^2 + m^2_{S}|S|^2  + \lambda A_{\lambda} \hat{H_u} \hat{H_d} \hat{S} + \frac{1}{3} \kappa A_{\kappa}\hat{S}^3, 
\end{equation}
where $A_{\lambda}$ and $A_{\kappa}$ are free parameters that take positive and negative values. After  EW Symmetry Breaking (EWSB) in a CP-conserving scenario, the Higgs fields acquire non-zero VEVs resulting in three CP-even neutral Higgs bosons $H_i, i = 1,2,3$, with $m_{H_1} < m_{H_2} < m_{H_3}$, two CP-odd Higgs bosons $A_i, i=1,2$, with $m_{A_1} < m_{A_2}$ and one pair of charged Higgs bosons ($H^\pm$). The presence of additional terms in the NMSSM Superpotential helps in controlling the masses of the Higgses while keeping the free parameters well within the perturbative limits. At tree level, the NMSSM Higgs sector is described by the six independent parameters: $\lambda$, $\kappa$, $\tan\beta$, $A_{\lambda}$, $A_{\kappa}$ and $\mu_{\rm eff}$. Depending on the mixing between the double and singlet Higgs fields, one can easily fit the observed SM-like Higgs boson of mass at 125 GeV with one of the CP-even neutral Higgses of this model.

The presence of a singlino significantly alters the phenomenology of the neutralino sector. For instance, the lightest neutralino can be a prime DM candidate having a dominant singlino component and thereby satisfying experimental constraints on DM relic density and limits from direct and indirect detection. The couplings of these singlino-like neutralinos to SM particles are often suppressed, allowing it to also evade detection at colliders easily while leading to a wider range of possible signatures for both collider searches and DM experiments in the years to come.

%%================================+%%
\section{Parameter Space Scan and Constraints}
%%================================+%%
In this section, we begin by presenting a detailed account of the computational tools and frameworks employed to systematically explore the NMSSM parameter space, emphasizing the methodology adopted in conducting the scans.
Following this, we provide a comprehensive discussion of the various constraints applied throughout the analysis, addressing in turn the theoretical requirements that ensure consistency of the model, as well as the experimental bounds derived from current collider data and other relevant measurements.
%%================================+%%
\subsection{Setup for DL Scanning of the NMSSM Parameter Space}
%%================================+%%
We start this subsection by describing the method used to test relevant constraints and obtain suitable regions of parameters.
The scan over the parameter space of the NMSSM utilizes 
the collection of tools
packaged into \texttt{NMSSMTools}.
This package encompasses the programs \texttt{NMHDECAY}~\cite{Ellwanger:2004xm,Ellwanger:2005dv},
\texttt{NMSPEC}~\cite{Ellwanger:2006rn},
\texttt{NMGMSB}~\cite{Ellwanger:2008py},
\texttt{NMHDECAY\_CPV}~\cite{Domingo:2015qaa},
\texttt{NMSDECAY}~\cite{Das:2011dg},
\texttt{micrOMEGAs}~\cite{Alguero:2023zol}
and \texttt{SModelS}~\cite{Alguero:2021dig}.
For faster calculation of an array of several parameters of the NMSSM,
we have created a custom interface in Python that takes care of running different instances of
{\tt NMSSMTools} in multiple CPUs and read their outputs into simple lists of parameter vectors.
This is particularly useful considering that our scan is performed
using the DL parameter scanning tool \texttt{DLScanner}~\cite{Hammad:2024tzz},
which has been implemented as a Python module.
This scanning tool uses Deep Neural Networks (DNNs) to learn and guess parameter regions
where a series of constraints may be met.
It does so by using either regression into a function that quantifies the distance to the best fit point
or classification of regions according to whether or not they follow all the constraints.
Moreover, it follows an iterative predict-learn loop where after results are predicted for new points, then 
the corrected results are used to keep training the network.
Additionally, it employs {\tt VEGAS}~\cite{Lepage:2020tgj} mappings to improve sampling of the parameter space
even when trying to identify comparably small regions inside a much larger space.

In this work, we have performed a scan over 12 parameters of the NMSSM, defined at the EW scale,
namely, $\tan\beta$, $\lambda$, $\kappa$, $A_\lambda$, $A_\kappa$,
$\mu_\text{eff}$, $M_1$, $M_2$, $M_3$, $A_t$, $M_{Q_3}$ and $M_{U_3}$ (see Ref.~\cite{Ellwanger:2009dp} for their definition.).
We start scanning over a large space in order to identify likely regions
before reducing the space.
The wide and narrow ranges used in the preliminary and the final scans, respectively,
are given in Table~\ref{tab:scanranges}.
The wide ranges of the preliminary scan are consistent with taking maximum upper values and minimum lower values
from the ranges scanned in Refs.~\cite{Ellwanger:2023zjc} and~\cite{Ellwanger:2024vvs},
except for $M_2$, $M_{Q_3}$ and $M_{U_3}$, which were allowed to run up to $10^4$~GeV.

\begin{table}[htb]
\begin{center}
\begin{tabular}{ccccc}
 \toprule
  & $\tan\beta$ & $\lambda$ & $\kappa$ & $A_\lambda$ \\
    wide & [1.97, 10.9] & [0.013, 0.687] & [0.0058, 0.391] & [$-5000$, 480] \\
    narrow & [3.2, 6.2] & [0.07, 0.42] & [0.05, 0.3] & [351, 834] \\
    \midrule
  & $A_\kappa$ & $\mu_\text{eff}$ & $M_1$ & $M_2$ \\
    wide & [$-621$, 362] & [$-244$, 291] & [178, 3000] & [304, 10000] \\
    narrow &  [$-300$, $-150$] & [120, 220] & [500, 3000] & [750, 10000] \\
    \midrule
  & $M_3$ & $A_t$ & $M_{Q_3}$ & $M_{U_3}$ \\
    & [423, 5000] & [$-5000$, 1288] & [272, 10000] & [570, 10000] \\
 \bottomrule
\end{tabular}
\end{center}
\caption{\label{tab:scanranges}
    Ranges scanned for the 12 parameters considered in this study.
    The word `wide' stands for the ranges used in the preliminary scan    while 
    the word `narrow' is used for the scan over the reduced parameter space.
    Parameters that do not indicate wide or narrow ranges
    receive the least effects from constraints and are therefore kept in the same range.
    Dimensionful parameters are given in GeV.
}
\end{table}

The scanning is done using the classification method in \texttt{DLScanner}.
By definition, classification requires a function that returns a
binary output that distinguishes between two classes:
in our case whether the points meets the conditions or not.
However, when testing many constraints, with wide parameter ranges
and scanning many parameters,
it is possible that the target region is several orders of magnitude
smaller than the scanned space.

This requires an approach that is flexible to
the possibility that only subsets of constraints may be satisfied
in the first few attempts,
while quantifying the closeness to the target region.

Therefore, instead of an approach purely based on classification,
we follow a hybrid approach based on iteratively sampling downwards
a penalty function.
To be more specific,
we define a penalty function based on constraints
that are difficult to satisfy at the same time,
if those constraints are not met for enough points,
we pick a large number of points with the smallest penalty function
and define them as the ``1'' class and the rest as the ``0'' class.
This, combined with the {\tt VEGAS} map attempting to sample the ``1'' class,
progressively reduces the minimum value found for the penalty function.
The type of penalty function we use is defined for each constraint
and is based on having 0 penalty when the set of parameters
gives a result within $\pm2\sigma$ and increases as results
move away from this range.
The penalty function for each constraint is defined as
\begin{equation}
    \label{eq:penalty}
    P_\text{constraint} (O^\text{th}, O^\text{exp}_{\pm 2\sigma}) = \begin{cases}
        0, &  O^\text{exp}_{-2\sigma} < O^\text{th} < O^\text{exp}_{+2\sigma}, \\
        |O^\text{exp}_{+2\sigma} -  O^\text{th}|^2, & O^\text{th} > O^\text{exp}_{+2\sigma}, \text{ if } O^\text{exp}_{+2\sigma} \text{ exists,}\\
        |O^\text{th} - O^\text{exp}_{-2\sigma}|^2, & O^\text{th} < O^\text{exp}_{-2\sigma}, \text{ if } O^\text{exp}_{-2\sigma} \text{ exists},
    \end{cases}
\end{equation}
where $O^\text{th}$ indicates the prediction obtained in our calculation and $O^\text{exp}$ stands for experimental values with $\pm2\sigma$ indicating
the $2\sigma$ limits reported by the experiment.
Note that, for constraints that only have an upper or lower bound, we can only make the comparison
against corresponding upper or lower bounds.

The penalties for the constraints are divided over their $1\sigma$ error
and then squared and summed over to make a global penalty for each point.
In the case of direct detection cross-sections for DM,
we divide their predicted values over the current limit
and take the base-10 logarithm of that value.
This stops direct detection from being the largest contribution to the total penalty.
Note that this penalty approach is only needed when some constraints are difficult
to satisfy at the same time.
In our case, these are relic density and direct detection of DM,
the masses for the three scalars (at $\sim 125$~GeV, $\sim 95$~GeV, $\sim 650$~GeV) 
and the Higgs reduced couplings.
Constraints that are easy to satisfy in large regions of parameter space
can be used to directly reject regions without applying any penalty.
Most notably, parameter combinations that happen to be unphysical
have to be outright rejected as well as combinations without a proper DM candidate.

For the running of the scan, we have used 140 steps in each phase of the scans (wide and narrow),
testing $10^7$ points through the DNN and selecting 10,000 of these 
that were predicted to be closer to the target space.
The network used was a fully connected DNN with two hidden layers
with 1000 neurons per layer trained for 1000 epochs.
The output layer uses a sigmoid activation function
while all the hidden layer nodes use a Rectified Linear Unit (ReLU) activation function.
For training of the network we use categorical cross-entropy for the loss function
with the Adaptive Moment Estimation (Adam) optimization algorithm.

As mentioned before, {\tt DLScanner} uses {\tt VEGAS} to create a map that transforms
a random distribution to another one following weights given by a data set.
This is done by training a {\tt VEGAS} map using 100 increments with all the available data during each step.
For this scan, we set the distribution of the $10^7$ test points
passed to the network as 75\% coming from a {\tt VEGAS} map and 25\% from a random uniform distribution.
The constraints used during the scan and their results are given in detail in the rest of this section.\footnote{
    
    The code used for the scan, example of inputs for \texttt{NMSSMTools}
    and their numerical results are available in the \texttt{DLScanner} repository via the following URL:\\
    \href{https://github.com/raalraan/DLScanner/tree/main/tests/examples/2508_13912}{https://github.com/raalraan/DLScanner/tree/main/tests/examples/2508\_13912}.
}

%%================================+%%
\subsection{Theoretical Constraints} 
%%================================+%%

Theoretical constraints such as perturbative unitarity, vacuum stability and perturbativity of the  quartic couplings are essential for the validity of the model. For the NMSSM, these requirements are particularly important. Specifically, the absence of a Landau singularity for the Yukawa couplings below the GUT scale imposes a strict constraint on the NMSSM-specific coupling, $\lambda$. This constraint limits the value of $\lambda$ to less than 0.7.
%{\textcolor{violet}{
For the minimum of the potential,
NMSSMTools checks that the physical
minimum is not a false minimum by comparing
against the value of the potential
when the scalar fields vanish.
%}}
We apply all of these constraints to the sampled points using the default implementation of {\tt NMSSMTools}.

These constraints, if not satisfied, automatically assign the corresponding point to the rejected class.
Only points that satisfy all theoretical constraints have their penalty function calculated as in Eq.~\eqref{eq:penalty}.

%{\color{blue} [AC: need to check with the code to get the list of theoretical constraints ...  ]}
%{\textcolor{violet}{
%[RR: From NMSSMTools source code,
%asking to check theoretical conditions means
%checking Laundau poles at the GUT scale
%and checking for false minima]
%}}

%%================================+%%
\subsection{Experimental Constraints} 
%%================================+%%
The list of experimental constraints that must be taken into account in our analysis is extensive and diverse, reflecting the wide range of collider searches and precision measurements relevant to the NMSSM framework. To present these in a clear and systematic manner, we organize the discussion into several subsections, each devoted to a particular class of constraints, thereby allowing us to highlight their individual roles and implications in shaping the viable parameter space.
%%================================+%%
\subsubsection{Higgs Measurements}
%%================================+%%

\begin{figure}[!htb]
    \includegraphics[width=1\linewidth]{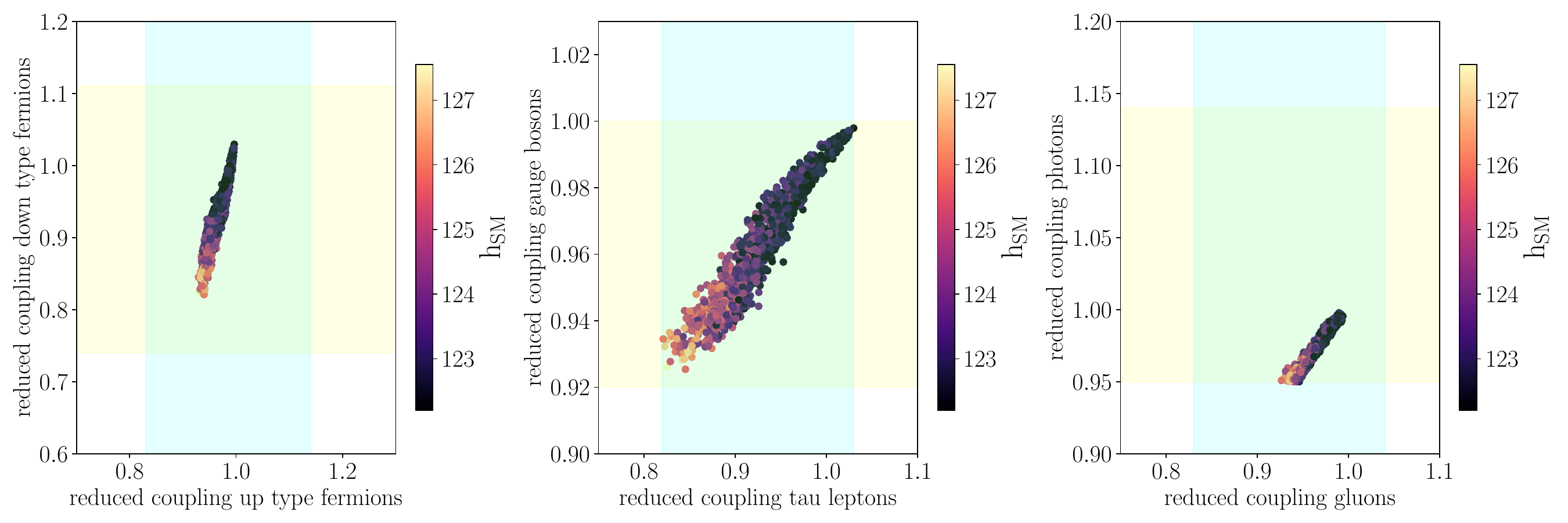}
    \caption{Reduced coupling correlations for the SM Higgs boson. The color bar represents the SM Higgs boson mass while the shaded areas represent the allowed region within $2\sigma$ away from the measured central values.}
    \label{fig:reduced_couplings}
\end{figure}

The discovery of a SM-like Higgs boson with a mass around 125 GeV by the ATLAS and CMS experiments at the LHC at CERN marks an important milestone in our quest to understand the dynamics of EWSB. Since its discovery, a major focus of research at the LHC has been to precisely measure the properties of this new particle and to verify whether it matches the predictions of the SM. So far no conclusive deviations from the SM expectations are observed, which in turn places strong constraints on any BSM parameter space. Therefore, every BSM set up with an extended scalar sector, either by modifying the couplings of the SM-like Higgs boson or by adding new BSM (pseudo)scalar states to the theory, must take into account the limits from the LHC as well as LEP and Tevatron data. To account for this large number of available searches and measurements, we use the dedicated software suite {\tt HiggsTools}~\cite{Bahl:2022igd} that combines the implementations of {\tt HiggsSignals}~\cite{Bechtle:2020uwn} and {\tt HiggsBounds}~\cite{Bechtle:2020pkv}, where the former checks against exclusion limits from direct searches for new Higgs particles while the latter evaluates the compatibility of the BSM scenario under consideration with the measurements of the 125 GeV Higgs boson. We scan the NMSSM parameter space and select those points where the second lightest CP-even Higgs boson ($H_2$) behaves like the SM-like Higgs boson with mass within a $2 \sigma$ from the measured value, with theoretical uncertainty of $\pm 3$ GeV. Although the $2\sigma$ mass uncertainty measured by ATLAS and CMS is  sub-GeV \cite{ATLAS:2015yey}, we consider a $\pm3$ GeV to account for the theoretical uncertainty that mostly originate from missing higher order loop corrections, different renormalization schemes and numerical approximations. The lightest ($H_1$) and heaviest ($H_3$) CP-even Higgs boson masses are set around 95 GeV and 650 GeV to accommodate the excesses observed by the LHC in those mass ranges. We discuss these excesses in detail later in this section.  The measurements around the 125 GeV Higgs boson are generally parametrized  in terms of the so-called `signal strengths' defined as follows: 
\begin{equation}
 \mu^i_j =  \frac{\sigma_i \times {{\rm BR}_j}}{{(\sigma_i \times {{\rm BR}_j})_{\rm SM}}}  =  c_i^2 \cdot \frac{{\rm BR}_j}{({{\rm BR}_j})_{\rm SM}},
\end{equation} 
where the reduced coupling $c^2_i$ equals the reduced
cross-section for the production mode $i$ and ${{\rm BR}_j}$ denotes
the corresponding Branching Ratio (BR) for the decay mode $j$.  
At tree level, these reduced couplings depend on $\tan\beta$ and Higgs mixing angle. However, loop-induced reduced couplings, involving $Z$'s, photons or gluons, may receive large contributions from other particles depending on their coupling with the Higgs boson. For all scanned points, we have calculated these reduced couplings, which are then used by {\tt HiggsTools} to verify the consistency with the available data. In fact, a dedicated $\chi^2$ analysis is performed using the available experimental measurements and theoretical predictions (here, NMSSM expectations) that include the correlations between the various measurements. Those points with $\Delta \chi^2$, defined as the difference of the calculated values of $\chi^2$ from the SM and NMSSM predictions, within a 2$\sigma$ interval off the central value, are considered for the subsequent analysis. This methodology helps us not only to probe the limits on the couplings relevant for the SM Higgs boson with mass close to 125 GeV but also to limit the possible amount of mixing between all three scalars, namely, the 95, 125 and 650 GeV CP-even Higgs bosons present in the BSM setup under consideration. In Figure \ref{fig:reduced_couplings}, we show the values of some of these reduced couplings of the SM Higgs boson. The color bar represents the SM Higgs boson mass while the shaded areas represent the allowed region within $2\sigma$ away from the central value. It is evident from the figure that all the scatter points are well within the allowed limits by the LHC data.  

%{\color{blue} [AC: Did we check the $H/A \to \tau^+\tau^-$ and $H^\pm \to \tau \nu$ constraints?]}

%%================================+%%
\subsubsection{DM Measurements}
%%================================+%%
%The cold DM problem stands as one of the most intriguing unresolved questions in Science.
Unraveling the true nature of DM remains one of the most profound challenges in particle physics. The constraints on the BSM parameter space coming from  DM searches are broadly classified into two categories: relic density constraints and limits from the  (in)direct searches. The DM relic density must satisfy the following limit as obtained from  Planck data~\cite{Planck:2018vyg}, 
\begin{equation}
 \Omega_{\rm CDM} h^2 =   0.120 \pm 0.001. \nonumber  
\end{equation}
%In our study, we assume that the LSP is the single contributor to the observed total DM relic density. We employ {\tt micrOMEGAs} to calculate the DM relic density for all the scanned data points and demand that all points must satisfy the $2\sigma$ upper bound of the observed relic density, i.e., 
%\begin{equation}
% \Omega_{\rm CDM} h^2 <   0.122. \nonumber
%\end{equation}

The neutral LSP is assumed to contribute to the relic density of the universe. However, we do not impose the requirement that it constitutes the entirety of DM, thus accommodating further contributions from physics at or above the EW scale. We employ the {\tt micrOMEGAs} package to compute the DM relic density for all data points and only those that satisfy the observed upper bound are retained, i.e., 
\begin{equation}
 \Omega_{\rm CDM} h^2 <   0.122. \nonumber
\end{equation}

We acknowledge that our choice for $\Omega_{\rm CDM} h^2$ might seem inconsistent with the standard practice of including a $10\%$ theoretical uncertainty around the central value. However, the requirement to explain the observed anomalies necessitates a highly specific spectrum: i.e., the LSP, Next-to-LSP (NLSP) and lightest chargino are nearly mass degenerate with the LSP having moderate higgsino-singlino mixing. This mass degeneracy significantly enhances both annihilation and co-annihilation processes, resulting in a highly suppressed relic density. Our finding is consistent with previous work (e.g., Ref. \cite{Ellwanger:2023zjc}),  which observed that relic densities are typically $O(10^{-3})$ or less when the LSP allows significant higgsino-singlino mixing.

%------------------------------
\begin{figure}[!h]
    \includegraphics[width=1\linewidth]{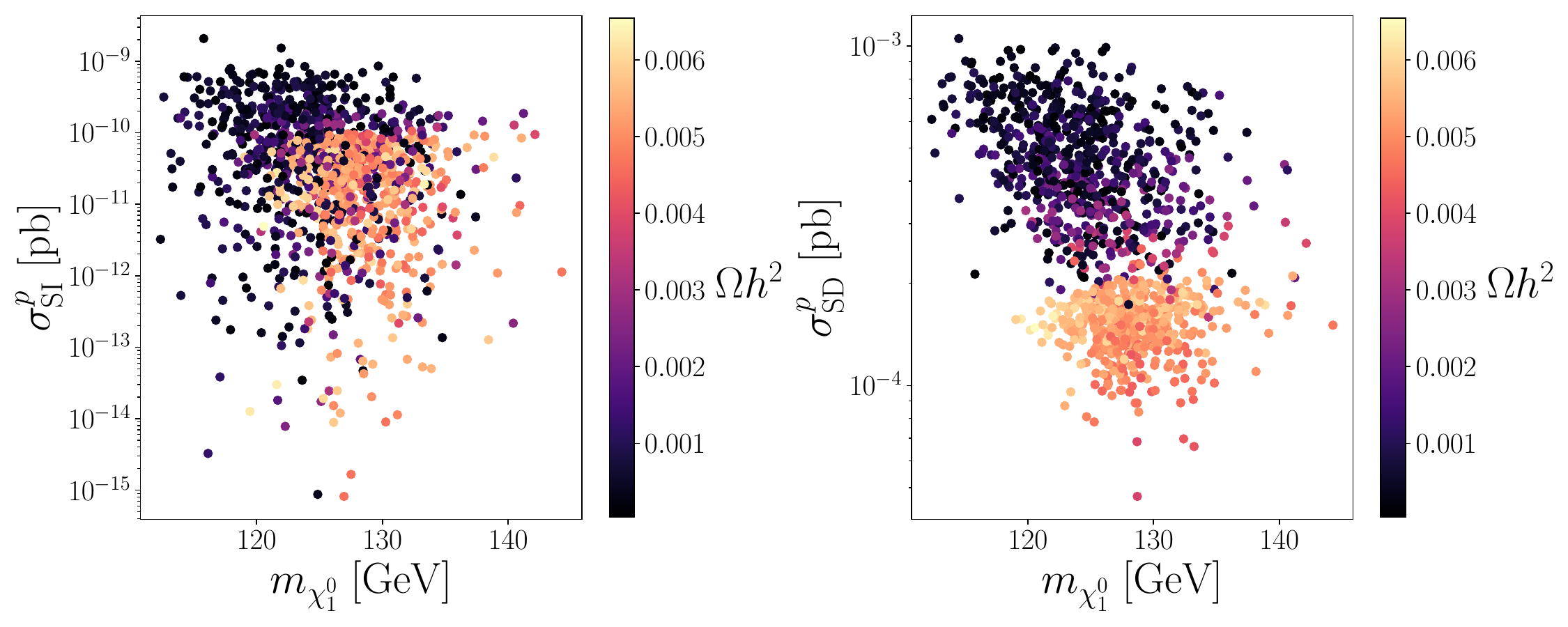}
    \caption{Left(Right): SI(SD) measurement versus the mass of the LSP. The color bar represents the relic abundance.
    \label{fig:relic-DD}}
\end{figure}

%===============================

Also, we take into account the constraints from direct DM searches by calculating the spin-independent (SI) and spin-dependent (SD) scattering cross-sections of the dark matter candidate, using {\tt micrOMEGAs}. In Figure~\ref{fig:relic-DD}, we present the SI (left) and SD (right) cross-sections scaled by  $\frac{\Omega_{\rm th} h^2}{\Omega_{\rm exp} h^2}$as a function of the LSP mass, with the color bar representing the corresponding relic density for the sampled points with  $\Omega_{\rm th} h^2$ being the calculated relic density and $\Omega_{\rm exp} h^2$ being the observed one. All points considered are consistent with both the observed relic density and the current limits from direct detection experiments, including LZ and XENON~\cite{LZ:2022lsv,XENON:2018voc}.

It is noteworthy that the interplay between the LSP composition and the extended Higgs sector in the NMSSM plays a crucial role in determining the scattering cross-sections. In fact, the mixing between the singlino and higgsino components in the LSP controls the DM scattering cross-section, primarily by offering a mechanism to suppress the interaction rate with nuclei in direct detection experiments. While a pure higgsino LSP generally has relatively large couplings to the Higgs and $Z$ bosons leading to large DM-nucleon scattering cross-sections that are often incompatible with current limits, a significant singlino component typically leads to suppressed SI interactions and becomes consistent with observations. When the LSP is a singlino-higgsino mixture, the overall coupling to the Higgs and $Z$ bosons is attenuated by the singlino mixture. This results in relatively smaller scattering cross-sections, allowing the mixed LSP to remain a viable DM candidate. It is the latter case, which helps to satisfy the current experimental bounds for all our sampled points, demonstrating that the NMSSM can accommodate viable DM candidates without conflicting with direct detection data.
%Points with a significant singlino component typically lead to suppressed SI interactions, while those with more Higgsino admixture can exhibit enhanced scattering rates. 
Moreover, these points are also within the projected sensitivities of upcoming experiments such as XENON-nT and the LZ upgrade~\cite{XENON:2020kmp,LZ:2018qzl}, emphasizing that a substantial portion of the currently allowed parameter space could be probed in the near future. 
Regarding DM indirect detection,
considering that our DM relic density tends to be of $\mathcal{O}(10^{-3})$,
as shown in Fig.~\ref{fig:relic-DD},
and with most averaged annihilation cross-sections times velocity typically below $10^{-25} \ \text{cm}^3 \text{s}^{-1}$,
we do not expect indirect detection measurements~\cite{Fermi-LAT:2013sme} to be a serious concern.

%%================================+%%
\subsubsection{Low Energy Observables}
%%================================+%%
In the NMSSM, precision measurements in the  \emph{B}-meson sector impose important constraints on the allowed parameter space. Rare flavor changing neutral current processes such as $B \to X_s \gamma$, $B_s \to \mu^+\mu^-$, and $B \to \tau \nu$, as well as mass differences in neutral meson  mixing like $\Delta M_{B_s}$, are particularly sensitive to new physics contributions. These observables receive loop corrections from charged Higgs bosons, charginos, stops, and other  superpartners, which can shift their predictions away from SM values.  Given the high precision of current experimental results from LHCb, Belle~II, and earlier  \emph{B}-factories, even small deviations can tightly constrain the NMSSM parameters,  making \emph{B}-physics an indispensable tool for probing its phenomenology in parallel  with collider and dark matter searches.

%The presence of light pseudoscalar and charged Higgs bosons naturally invites the constraints from flavor physics for the parameter space of our interest. {\textcolor{red}{SM: Ahmed/Raymundo, the previous sentence is out of nowhere, as we did not discuss $A_{1,2}$ and $H^\pm$ states anywhere, see AC's comments too... They should be appropriately introduced, including mentioning constraints on their properties from precision data (i.e., $S,T,U$)}}
%{\color{violet}[RR: I am leaving this question to whomever made the \texttt{FlavorKit} analysis]}
%We calculate several flavor observables for each scan point using {\tt FlavorKit}~\cite{Porod:2014xia} {\textcolor{green}{[AH: Do we really use FlavorKit? The files I have use NMSSMTools only!]}}.

The following two  observables are found to play a crucial role,  $B\to S\gamma$ and $B^\pm\to \tau^\pm\nu_\tau$, in constraining the values of $m_H^\pm$ and $\tan\beta$. We consider the experimental bounds as reported in~\cite{HFLAV:2019otj} and allow for a  $2\sigma$ deviation from their central value. 
\begin{align}
\text{BR}(B \to X_s \gamma)_{E_\gamma \ge 1.6~\mathrm{GeV}} & = (3.32 \pm 0.15) \times 10^{-4}\,, \\
\text{BR}(B^\pm \to \tau^\pm \nu_\tau) & = (1.11 \pm 0.165) \times 10^{-4}\,.
\end{align}

\begin{figure}[!htb]
    \includegraphics[width=1.\linewidth]{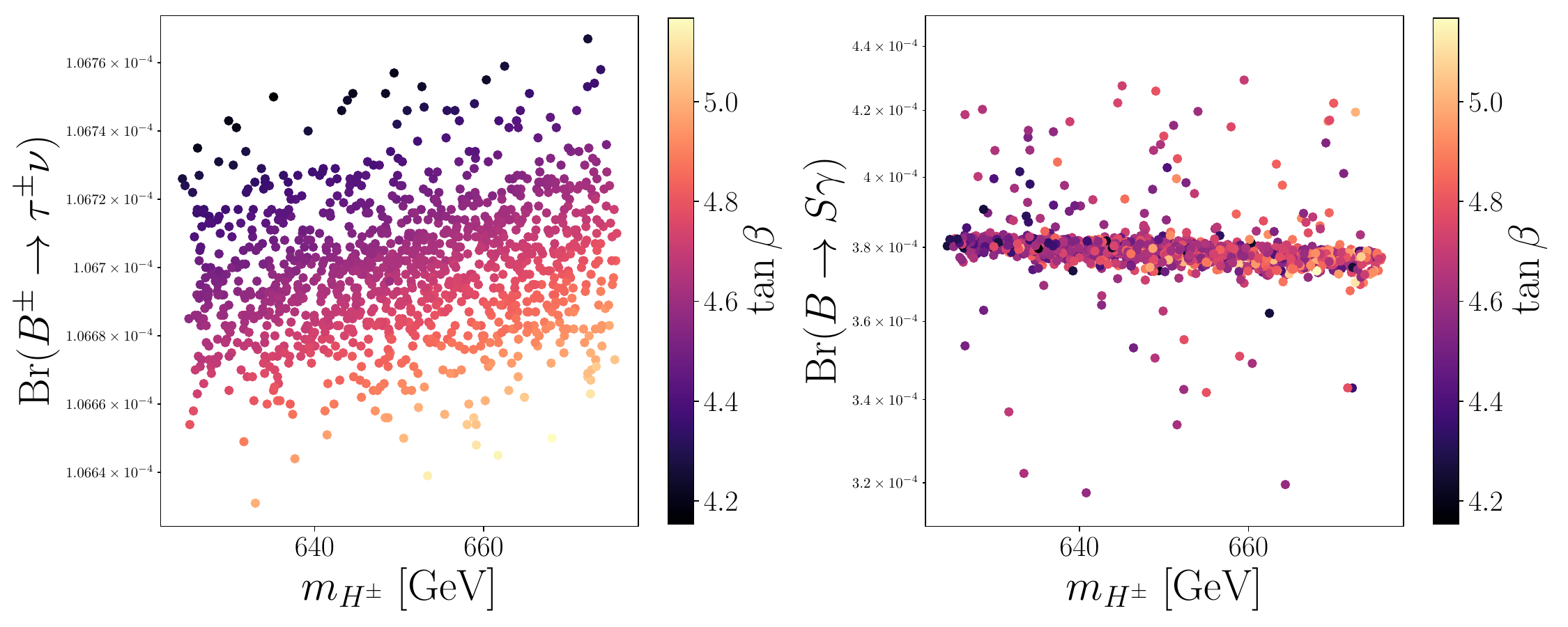}
    \caption{Left(Right): NMSSM allowed scan points following $B^\pm \to \tau^\pm\nu$($B\to S\gamma$) constraints. The color bar represents $\tan\beta$ value. }
    \label{fig:low_obs}
\end{figure}

In the SM, the dominant contribution to BR$(B\to S\gamma)$ comes from the $t-W^\pm$ loop, almost saturating the measured value. However, in SUSY theories, significant contributions may arise if the models include light SUSY particles. In the NMSSM framework, the  $\tilde{t_1} - \chi_1^\pm$ loops can provide large contributions to the above-mentioned decay mode. Similarly, decays of  $B$-mesons to final states with $\tau$-leptons are also sensitive to BSM effects, especially those involving a charged Higgs boson. Of course, all of these effects are mostly sensitive to the low $\tan\beta$ region. In Figure \ref{fig:low_obs}, we show the points allowed by these two constraints with the color mapped against values of $\tan\beta$. As expected, the allowed points have $m_H^\pm$ greater than 600 GeV with $\tan\beta$ values less than 6, thus significantly minimizing the effects of new physics. The top quarks are decoupled and, therefore, the corresponding loop contribution to these flavor decays is highly suppressed. Furthermore, $\text{BR}(B_s \to \mu^+ \mu^-)$ has been verified and found to be consistent with the current experimental bounds of $(3.1 \pm 0.7) \times 10^{-9}$.

We also apply the constraints from EW Precision Observables (EWPOs), a powerful way to scrutinize the SM at the highest precision. We use \texttt{ NMSSMCALC}~\cite{Baglio:2013iia}, a publicly available code to calculate the oblique parameters (namely S, T, and U) for the NMSSM, and demand that the estimated values of S,T and U for the sampled points are within the $2\sigma$ of the best fit values provided by the {\tt GFITTER}~\cite{Baak:2014ora}, assuming the reference mass of the Higgs boson and Top quark at 125 GeV and 173 GeV, respectively. 
\begin{equation}
   S=0.05\pm 0.11, \quad T=0.09\pm 0.13, \quad {\rm and} \quad U=0.01\pm 0.11. \label{eq:STU-numbers}
\end{equation}

%%================================+%%
\subsubsection{Collider constraints}
%%================================+%%
We now focus on the direct collider constraints on the NMSSM parameter space. A variety of experimental bounds on the NMSSM have been explored in the literature, although different studies have emphasized different sets of constraints. For example, Ref.~\cite{Barger:2006dh} primarily examined LEP II limits with a particular emphasis on scenarios with small  $\lambda$. To account for constraints from  direct collider searches of (s)particles we employ the \texttt{SModelS} framework~\cite{Ambrogi:2017neo}, integrated within the \texttt{NMSSMTools} package.

\texttt{SModelS} is a publicly available tool designed to interpret and constrain theories beyond the Standard Model using existing results from the LHC. Instead of performing a full simulation of complete new-physics models, \texttt{SModelS} decomposes their predicted collider signatures into simplified topologies defined solely by particle masses, production cross-sections, and decay chains. These simplified signatures are then confronted with the extensive database of experimental limits from ATLAS and CMS. This approach allows one to efficiently determine whether a given BSM scenario is excluded or constrained by current searches without the need for dedicated recasting of each analysis. 

Although in the scanned points squarks and gluinos are pushed to the TeV mass range, \texttt{SModelS} indicates that the strongest constraints still arise from searches for squarks and gluinos in final states with jets and missing transverse energy. In particular, the most stringent exclusions are derived from ATLAS~\cite{ATLAS:2020syg,ATLAS:2017mjy} and CMS~\cite{CMS:2019zmd,CMS:2021beq}analyses targeting jets plus missing-energy signatures.
%%================================+%%
\section{Anomalies}
%%================================+%%
In this section, we present a detailed discussion of the anomalous datasets that are employed to further probe the NMSSM parameter space. The purpose of this analysis is to investigate whether the NMSSM framework can provide a consistent and viable explanation for these anomalies, thereby offering potential insights into its phenomenological relevance. For clarity and completeness, each anomaly will be examined separately and in turn, with particular emphasis on its characteristics, the methodology of its incorporation into our study and the implications it carries for the viability of the model.

%%================================+%%
%%================================+%%
%%================================+%%
\subsection{The 95 GeV  Measurement}
%%================================+%% 

Both the CMS and ATLAS collaborations at the LHC have independently observed hints of a new particle with a mass around 95~GeV. In fact, the CMS experiment has consistently seen a slight excess in this mass range, first at 97~GeV with about $2\sigma$ significance and later at 95.4~GeV with similar significance in the diphoton final state in proton-proton collisions at 13 TeV~\cite{CMS:2015ocq,CMS:2018cyk,CMS:2024yhz}. The ATLAS experiment has also seen a similar, albeit smaller, excess of $1.7\sigma$ at 95~GeV in the same diphoton mode~\cite{ATLAS:2024bjr}. Interestingly, also the ALEPH, DELPHI, L3 and OPAL (ADLO) experiments at the LEP collider have found a small excess at a mass of 98~GeV decaying to $b\bar{b}$~\cite{LEP:2003ing}. 
%{\textcolor{violet}{[RR: It is not clear to me what ``corresponding decaying'' means in here. Should this be ``corresponding to $b\bar{b}$ decay''?]}}.
Given the large uncertainties in the $m_{b\bar b}$ resolution,  the oldest observations seem consistent with the recent ones. Although none of these findings are definitive enough, the consistent appearance of this signal across different experiments is interesting on its own. 

All of the above-mentioned observations are best explained in terms of the aforementioned signal strength variables. We consider two such observables, namely, $\mu_{\gamma\gamma}$ and  $\mu_{b\bar b}$, defined as follows: 
\begin{equation}
    \mu_{\gamma\gamma} = \frac{\sigma(g g \to h_{95} \to \gamma\gamma)}{\sigma(g g \to h^{\rm SM}_{95} \to \gamma\gamma)}\,, \hspace{6mm}  
    \mu_{b\bar b} = \frac{\sigma(e^+e^-\to Z h_{95} \to Z b\bar b)}{\sigma(e^+e^- \to Z h^{\rm SM}_{95} \to Z b\bar b)} 
    \nonumber,
\end{equation}
where $h_{95}$ denotes the new resonance of mass around 95~GeV and $h^{\rm SM}_{95}$ denotes the same but with SM-like couplings. The gluon-fusion process is considered as the production mode of the 95~GeV particle to analyze the $\gamma\gamma$ mode, while for the $b\bar b$ mode it is produced in association with a $Z$ gauge boson.  The total cross-section for each point in the scan is calculated using {\tt MadGraph}~\cite{Alwall:2014hca}. Following the combined analysis of ATLAS and CMS, we use $ \mu_{\gamma\gamma} = 0.24^{+0.09}_{-0.08}$~\cite{Biekotter:2023oen} while for $\mu_{b\bar b}$ we use the ADLO combined limit from LEP  $\mu_{b\bar b} = 0.117 \pm 0.057$~\cite{Cao:2016uwt}. We demand that the signal strengths calculated for the sampled points must satisfy these limits within a $2\sigma$ range of the central value.   
%%================================+%%
\begin{figure}[!htb]
    \includegraphics[width=1.\linewidth]{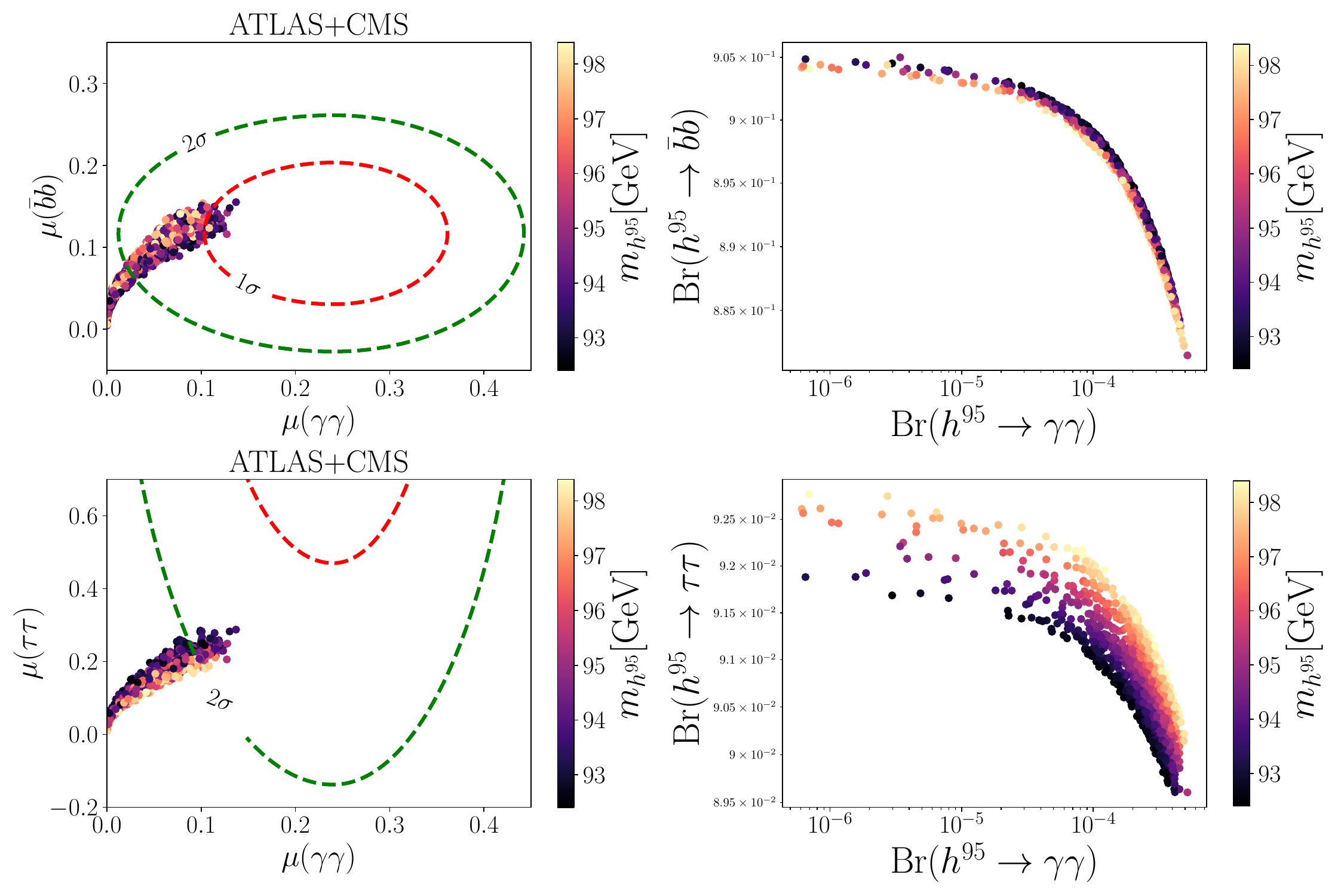}
    \caption{{{Distribution of points in the 2D plane of relevant signal strengths (left) and BRs (right) in order to explain the deviation at 95~GeV. The color gauge represents the $h_{95}$ mass.}}}
    \label{fig:signal_strength}
\end{figure}

Interestingly, CMS has also observed an excess of local significance of $2.7\sigma$ around 95~GeV in the search for a light neutral (pseudo)scalar boson in proton-proton collisions at the LHC with $\sqrt s = 13$ TeV in the di-tau channel~\cite{CMS:2022goy}. We also consider this mild but significant excess by calculating the corresponding signal strength variable as follows: 
\begin{equation}
    \mu_{\tau^+\tau^-} = \frac{\sigma(g g \to h_{95} \to \tau^+ \tau^-)}{\sigma(g g \to h^{\rm SM}_{95} \to \tau^+\tau^-)} \,.
    \nonumber
\end{equation}
We thus demand that the points must further satisfy $\mu_{\tau^+\tau^-} = 1.38 ^{+0.69}_{-0.55}$~\cite{CMS:2022goy}. Figure \ref{fig:signal_strength} displays the distributions of the various signal strength variables together with the relevant branching ratios of the 95 GeV resonance considered in this study. As can be seen, the majority of the scanned points lie well within the $2\sigma$ contour in the two-dimensional plane spanned by the signal strength variables.

In Section \ref{sec:results}, we present the explicit parameter space corresponding to points that remain within the $2\sigma$ level when all anomalies and constraints are simultaneously taken into account.

%%================================+%%
\subsection{The EWino Measurement}
%%================================+%%
We now turn our attention to another interesting observation made by the ATLAS and CMS collaborations at the LHC. Interestingly, both experiments have observed mild deviation (only at the 1-2$\sigma$ level) of events over the SM expectation corresponding to the final states with multiple leptons and large $\slashed{E}_T$. Specifically, these events correspond to chargino--neutralino pair production processes where chargino and neutralino decays proceed via an off-shell $W^\pm,Z$ boson leading to soft leptons~\cite{ATLAS:2019lng,CMS:2021edw,ATLAS:2021moa,CMS:2024gyw}. The simplified SUSY models considered in these studies are of two types, wino/bino and higgsino scenarios. In the wino/bino scenario, the $\tilde{\chi}^0_1$ (i.e., the LSP) is considered to be purely bino whereas $\tilde{\chi}^0_2$ and $\tilde{\chi}^{\pm}_1$ are taken purely wino and degenerate in mass. The simplified higgsino model assumes $\tilde{\chi}^0_1$, $\tilde{\chi}^0_2$ and $\tilde{\chi}^{\pm}_1$ to be higgsino-like states with compressed mass spectrum with $\tilde{\chi}^{\pm}_1$ lying approximately halfway between the two light neutralinos. 
In the wino/bino scenario, the mild deviation as reported in the observed versus expected no of events appears when the mass difference $\Delta m = m_{\tilde{\chi}^0_2} - m_{\tilde{\chi}^0_1} \sim$ 10--30~GeV when $\tilde{\chi}^0_2$ and $\tilde{\chi}^{\pm}_1$ decays to the LSP plus an off-shell $W^\pm,Z$ boson. In the higgsino scenario, the deviation is observed in the range $\Delta m \sim $ 10--30~GeV with the most pronounced effect at $\Delta m \sim 30$~GeV for the ATLAS collaboration. The CMS collaboration has also reported a similar deviation of events observed in SUSY searches with soft multi-lepton plus $\slashed{E}_T$ final state. Although CMS finds a mild deviation of events in the range $\Delta m \sim $ 8--30~GeV for the higgsino case, an excess of events is observed in the region of $\Delta m =$ 20--50~GeV for the wino/bino scenario. Considering the mass resolution, both deviations observed by the two collaborations are compatible with each other, with CMS finding deviations more on the higher mass differences.
Interestingly, similar deviations have also been observed in mono-jet searches around the same mass region~\cite{CMS:2021far,ATLAS:2021kxv}.

\begin{figure}[!ht]
    \centering
    \includegraphics[width=1\linewidth]{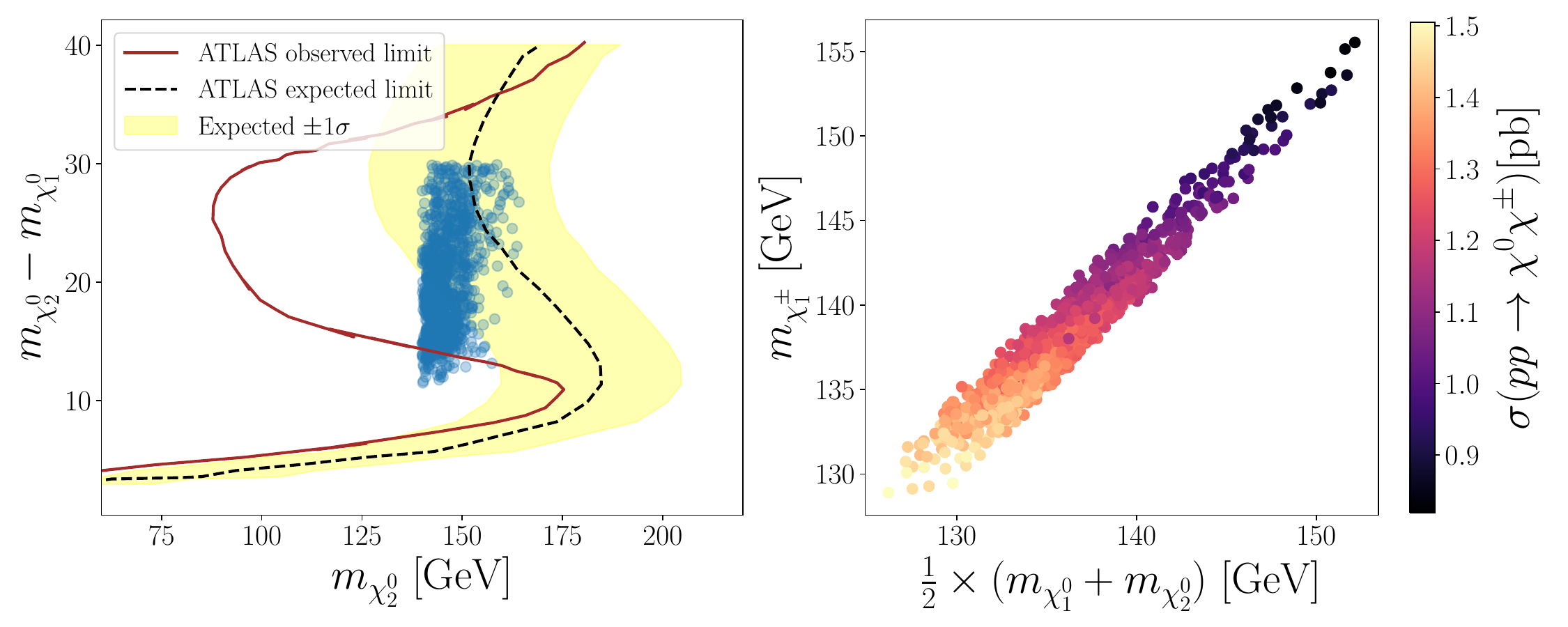}
    \caption{Left: Expected and observed ATLAS limits with scanned points overlaid in blue, which are demonstrating clear NMSSM compatibility with the EWino anomaly. ATLAS data are extracted from Figure 16 of~\cite{ATLAS:2021moa}. Right: cross-section for chargino--neutralino pair production.}
    \label{fig:eweakino1}
\end{figure}

Motivated by Ref.\cite{Ellwanger:2024vvs}, in our analysis, we study the higgsino model\footnote{While the study in \cite{Ellwanger:2024vvs} utilizes a singlino-like LSP to simultaneously explain the EWino search observations at the LHC alongside the $95\text{ GeV}$ anomaly in $b\bar{b}$ (LEP) and the $\gamma\gamma$ ($\text{LHC}$) one within the $\text{NMSSM}$, our approach is more comprehensive. In fact, we aim to explain these three observations plus a number of other recent anomalies. As a consequence, the resulting parameter space that successfully explains all these observations requires an $\text{LSP}$ that is predominantly higgsino-like with a moderate singlino admixture, as detailed later.} scenario in the NMSSM and consider a conservative limit $\Delta m =$ 10--30~GeV to identify the correct region of the parameter space. Figure \ref{fig:eweakino1} illustrates the distribution of sampled points that satisfy the constraints discussed above. The left panel shows the exclusion limits in the $m_{\tilde{\chi}^0_2} - \Delta m$ plane, while the right panel depicts the chargino--neutralino pair-production cross-section in the $m_{\tilde{\chi}^0_2} - m_{\tilde{\chi}^{\pm}_1}$ plane. The phase space regions providing the correct cross-section have both the neutralinos and lightest chargino masses below 250~GeV, as pointed out in other studies~\cite{Chakraborti:2024pdn}. The production cross-section for the scanned points was computed using {\texttt{MadGraph}.

For completeness, in Figure \ref{fig:eweakino2}, we show the amount of higgsino and singlino components in the LSP and NLSP in terms of the elements of the neutralino mixing matrix $\mathcal Z$, especially the higgsino-singlino elements ($ZNH$ and $ZNS$) with the wino/bino elements ($ZNB = ZNW$) vanishingly small. Note that the NLSP has been found to be of pure higgsino nature in order to be consistent with the observed measurement.

\begin{figure}[!ht]
    \centering
    \includegraphics[width=1\linewidth]{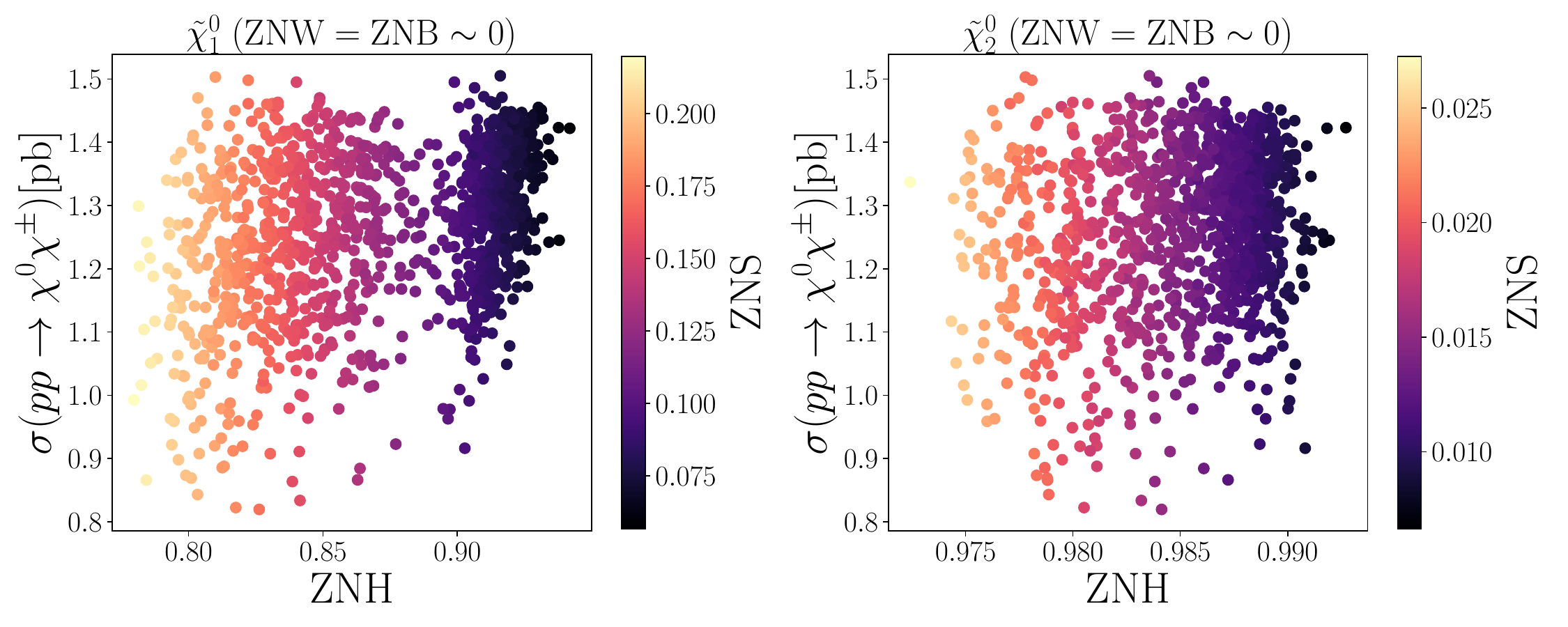}
    \caption{Plots to identify the LSP and NLSP types. The NLSP is a pure higgsino, which matches the experimental measurement.}
    \label{fig:eweakino2}
\end{figure}

%%================================+%%
%%================================+%%
\subsection{The 650~GeV  Measurement}
%%================================+%%
%%================================+%%
A search for the resonant production of a Standard Model Higgs boson in association with a new scalar particle, denoted as $Y$, via a heavy resonance $X$, observed a local excess of $3.8\sigma$ in the final state consisting of two photons (from the decay of the $h_{\rm SM}$) and two $b$-quarks (from the decay of $Y$)~\cite{CMS:2023boe}. The $95\%$ confidence level (CL) upper limit on the product of the production cross-section and branching ratio is approximately 0.8~fb. However, as reported in the experimental analysis, the best fit value of the cross-section with the branching fraction in the largest excess is $0.35^{+0.17}_{-0.13}$~fb. Therefore, the anomaly in the $bb\gamma\gamma$ channel can be fulfilled if the total cross-section lies within the best fit value. This excess was found for a strong resonance $X$ with $M_X \sim 650$~GeV and $M_Y \sim 90 - 100$~GeV. A previous study by the CMS Collaboration searched for a heavy resonance $X$ (with 240~GeV $< M_X < 3$TeV) decaying into a SM Higgs boson ($h_{\rm SM}$) and a lighter scalar $h_s$ (with 60GeV $< m_{h_s} < 2.8$~TeV). In this search, the $h_{\rm SM}$ decays into a pair of $\tau$ leptons, while the lighter scalar $h_s$ decays into a pair of $b$ quarks~\cite{CMS:2021yci}. No significant excess was observed, leading to model-independent 95\% CL upper limits on the product of the production cross-section and BRs for a wide range of $X$ and $Y\equiv h_s$ masses ranging between 125~fb (for $m_X$ = 240~GeV) to 2.7~fb (for $m_X$ = 1000~GeV).  While these observations are intriguing, the statistical significance is modest, with local estimates around $2 - 3 \sigma$, indicating that these excesses could well be due to fluctuations.

%------------------------
\begin{figure}[!th]
    \includegraphics[width=1.\linewidth]{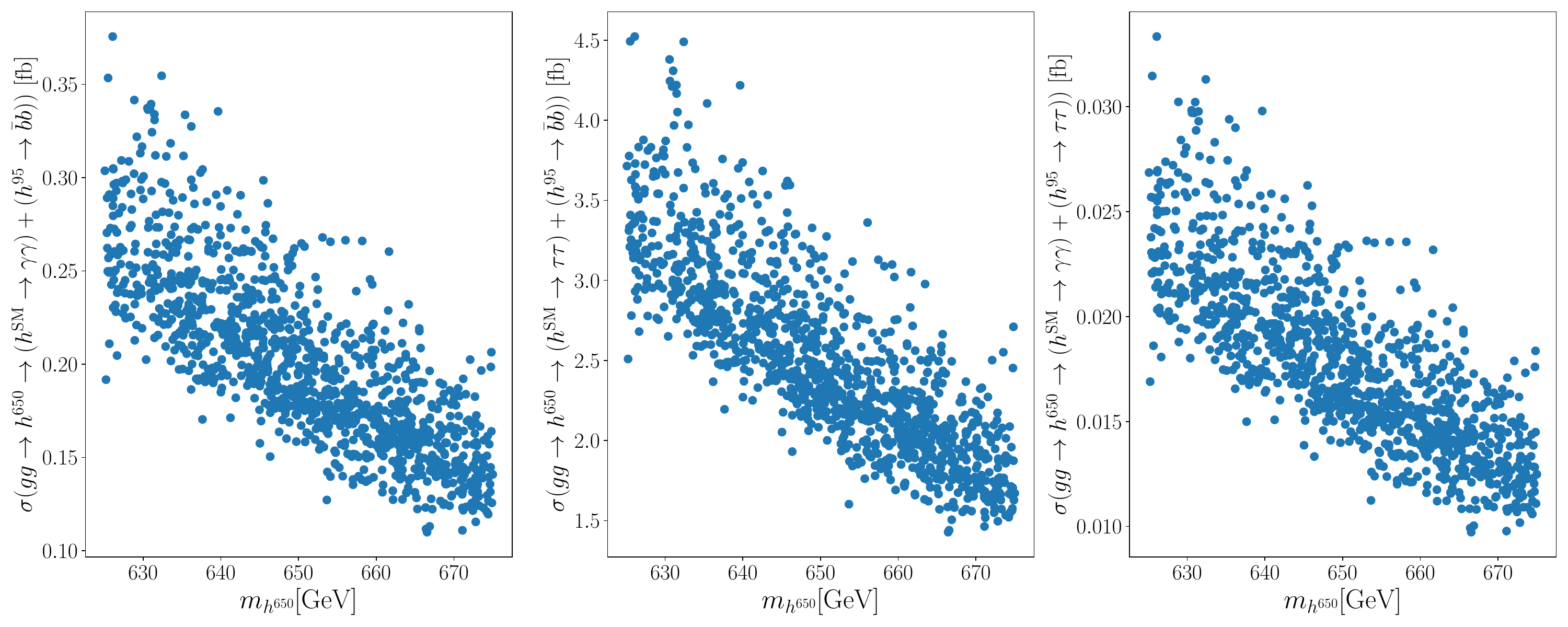}
    \caption{The distribution of $\sigma \times {\rm BR}$ values as a function of the $h^{650}(\equiv {H_3}\equiv X)$ mass for all the sampled points associated to the excess of events observed around 650~GeV.
    All the points in the left pane are inside the $\pm2\sigma$ range of $0.35^{+0.17}_{-0.13}$~fb,
    while for the points in the middle and right panes
    only upper bounds are set which are far outside the shown vertical ranges (see text.)
    }
    \label{fig:650}
\end{figure}
Complementing these searches, CMS recently performed a dedicated analysis targeting heavy resonances ($X$) decaying into the Standard Model Higgs boson ($h_{\rm SM}$) and a lighter scalar ($h_s$), with the characteristic final state $\gamma\gamma\tau^+\tau^-$~\cite{CMS:2025tqi}. This channel is particularly interesting as it probes scenarios with $m_{h_s}\approx 95$GeV, thereby directly testing one of the most intriguing excesses reported in recent years. The analysis sets stringent upper limits on the production cross-section times branching ratio for $X \to h_{\rm SM}h_s \to \gamma\gamma\tau^+\tau^-$, ranging from 0.69 to 15~fb in the low $X$-mass region. When combined with previous searches in complementary final states, such as $b\bar b\tau^+\tau^-$ and $\gamma\gamma b\bar b$, these results provide a powerful handle on probing the extended Higgs sector of the NMSSM.

We have incorporated all these experimental results to constrain the parameter space relevant to our study. As shown in Figure~\ref{fig:650}, the combination of limits from different final states significantly reduces the viable regions of parameter space, particularly those predicting sizable branching fractions into $b\bar b\tau^+\tau^-$. This highlights the strong interplay between different Higgs decay channels in shaping the phenomenology of the NMSSM and underlines the importance of multi-channel searches in covering scenarios where one individual signature might otherwise evade detection.
%%================================+%%
%%================================+%%
\subsection{The $(g-2)_\mu$ Measurement}
%%================================+%%
The anomalous magnetic moment of the muon, $a_\mu \equiv (g - 2)_\mu/2$, ranks among the most precisely determined quantities in physics, both experimentally and theoretically. This observable plays a pivotal role, as the persistent deviation between its measured value and the SM prediction could offer vital clues to the existence of BSM physics. The Fermilab Muon $g - 2$ experiment (E989) has reported the results from its Run-1 dataset, collected in 2018~\cite{Muong-2:2021ojo,Muong-2:2021ovs,Muong-2:2021xzz}, which show excellent agreement with the earlier E821 measurement performed at BNL~\cite{Muong-2:2006rrc}. The mild deviation between the experimental measurements and the SM prediction~\cite{Aoyama:2020ynm} is 
\begin{equation}
   a_\mu^{\rm{exp}} - a_\mu^{\rm{\rm SM}} =\Delta a_\mu = \left(24.9\pm 4.8\right) \times 10^{-10}\,,
\end{equation}
corresponding to a $5\sigma$ deviation. However, this result does not incorporate the lattice calculation for the leading order hadronic vacuum polarization contribution~\cite{Borsanyi:2020mff}. Inclusion of this theory prediction would yield a slightly higher value for $\Delta a_\mu$ leading to a smaller deviation from the experimental result of $< 2\sigma$. This leads to a difference of~\cite{Aliberti:2025beg}
\begin{equation}
\label{eq:deltamunew}
    a_\mu^{\rm{exp}} - a_\mu^{\rm{\rm SM}} =\left(38\pm 63\right) \times 10^{-11}\,.
\end{equation}
\begin{figure}[!h]
    \includegraphics[width=1.\linewidth]{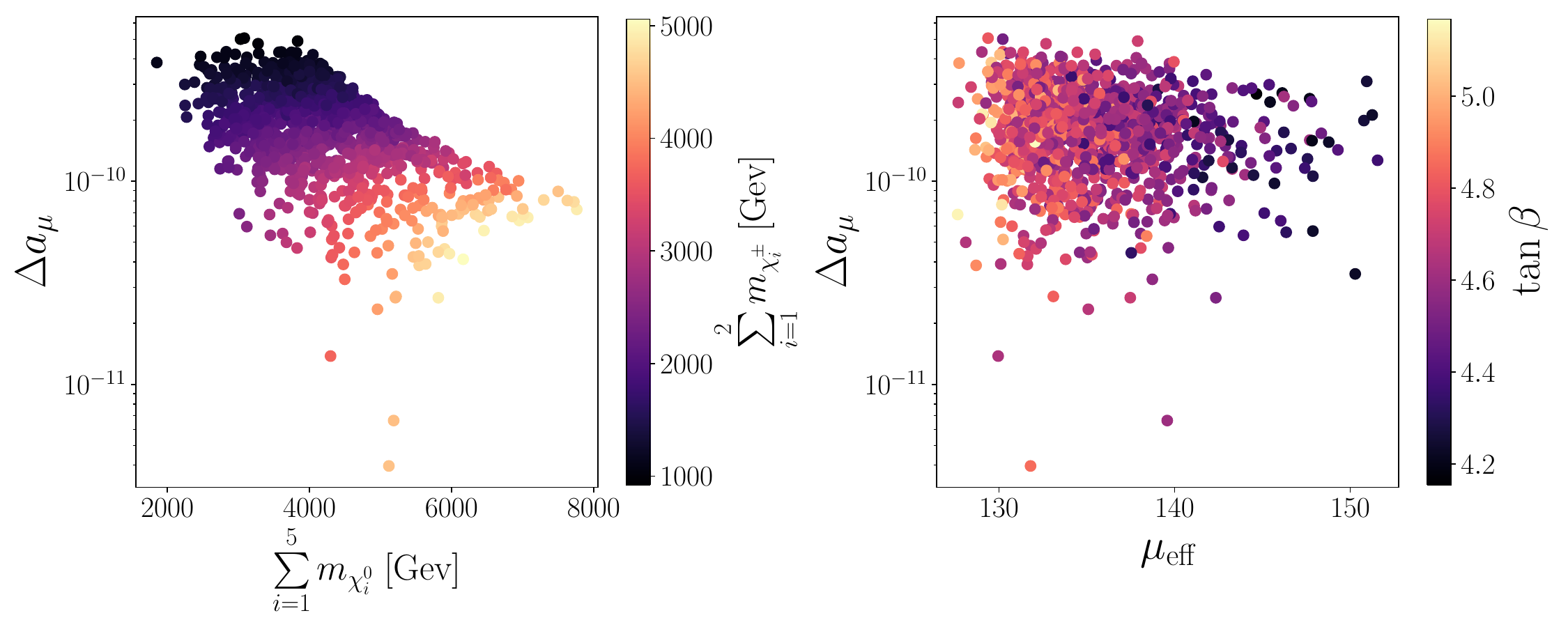}
    \caption{Left: Muon anomalous magnetic moment for the selected points plotted against the sum of neutralino masses on the $x$-axis, with the color scale indicating the sum of chargino masses. Right: Muon anomalous magnetic moment for the selected points plotted against $\mu_{\rm eff}$ on the $x$-axis, with the color scale indicating $\tan\beta$.}
    \label{fig:g2}
\end{figure}

In the NMSSM, the dominant one-loop contributions to the anomalous magnetic moment of the muon, $a_\mu$, arise from loops involving chargino--sneutrino or neutralino--smuon pairs. These contributions grow approximately linearly with $\tan\beta$ and are positive when the effective higgsino mass parameter, $\mu_{\rm eff}$, is positive. Figure \ref{fig:g2} presents the corresponding values of $\Delta a_\mu$ for the points selected in our scans. The left panel shows $\Delta a_\mu$ as a function of the sum of neutralino masses, with the color bar indicating the sum of the chargino masses. This plot demonstrates a clear trend of increasing $\Delta a_\mu$ for smaller values of either the chargino or neutralino masses. The right panel displays the same information in the plane of $\Delta a_\mu$ versus $\mu_{\rm eff}$, with the color bar representing $\tan\beta$, highlighting the dependence on these parameters. Overall, these results indicate that the selected points are consistent with the most recent measurement of the muon anomalous magnetic moment, as given in Equation~\ref{eq:deltamunew}, thereby providing further support for the viability of the NMSSM parameter space under consideration.
%%================================+%%

\section{Our Predictions}
In the spirit of not only producing NMSSM `postdictions', i.e., of explaining existing anomalous data {\sl a posteriori}, but also `predictions', i.e., pointing to {\sl new} datasets that will reveal themselves also anomalous if investigated over the viable parameter space, we chose among the latter two collider signatures of DM.  

%%%%%%
\subsection{Mono-$H$ and -$Z$ Signals}
%%================================+%%
%%================================+%%
In this subsection, we investigate the sensitivity of the HL-LHC to mono-$X$ signatures, focusing on BPs selected from the optimized NMSSM parameter space. Specifically, we study the production processes
\[
pp \to \tilde{\chi}_1^0 \tilde{\chi}_1^0 H \quad \text{and} \quad pp \to \tilde{\chi}_1^0 \tilde{\chi}_1^0 Z,
\]
where the lightest neutralinos $( \tilde{\chi}_1^0 )$ are weakly interacting massive particles  that escape detection, leading to $\slashed{E}_T$. These channels are particularly relevant for probing EWino DM at the HL-LHC.

(For studies of mono-$X$ signatures at the LHC, with $X=H,Z$, see Refs.~\cite{Bell:2012rg,Carpenter:2013xra,Petrov:2013nia,Berlin:2014cfa,Mattelaer:2015haa,No:2015xqa,Neubert:2015fka,Brivio:2015kia,Basirnia:2016szw,Arina:2016cqj,Barman:2016kgt,Belyaev:2016lok,Liew:2016oon,VanDong:2020nwb,Arcadi:2020gge,Adhikary:2020ujn,Dutta:2021ekc,Bahl:2021xmy,CMS:2024zqs,Arcadi:2024mli}.)
In principle, these signatures can lead to additional anomalies in future measurements that align with the current anomalies mentioned above. For this purpose, we employ a multi-modal network, that integrates graph-based DL layers and MLP layers, to analyze these signatures and enhance discrimination between signal and background. This approach integrates both high-level physics observables and low-level detector features, offering improved performance compared to conventional cut-based or shallow Machine Learning (ML) methods.

Although mono-jet channels have been widely studied and offer complementary sensitivity to DM and EWino searches~\cite{Agin:2025vgn,Fuks:2024qdt,Agin:2023yoq}, our focus here is on the mono-$H$ and -$Z$ channels, which provide distinctive kinematic features and are sensitive to different regions of the NMSSM parameter space. Feynman diagrams corresponding to the leading-order contributions to these channels are shown in Figure~\ref{fig:feynman}, with the upper row depicting the mono-$Z$ processes and the lower row representing mono-$H$ production.

\begin{figure}[!th]
    \centering
    \includegraphics[width=0.99\linewidth]{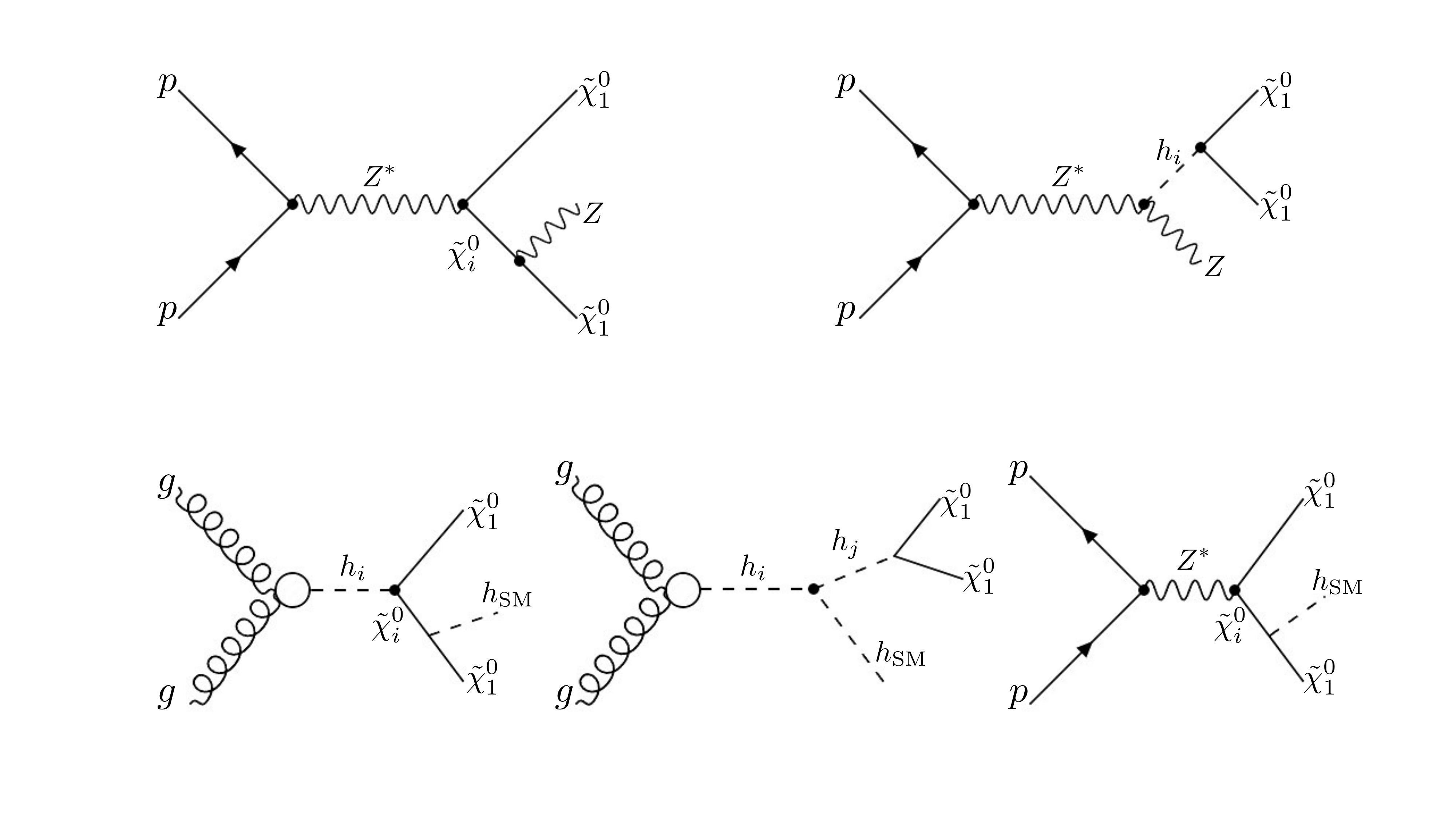}
    \caption{Feynman diagrams illustrating the leading contributing channels. The top row shows mono-$Z$ processes, while the bottom row displays mono-$H$ processes.}
    \label{fig:feynman}
\end{figure}

To quantify the expected signal rates, we compute the production cross-sections at the HL-LHC for each of the mono-$X$ channels. Figure~\ref{fig:sigma_mono}, left panel, shows the total cross-sections as a function of the sum of the masses of the lightest and next-to-lightest neutralinos. The mono-jet, mono-$Z$ and mono-$H$ channels are shown in blue, orange and green, respectively. The red and black markers indicate the specific BPs selected for the DL analysis in the mono-$Z$ and mono-$H$ channels, respectively. The right panel displays the cross-sections as a function of the NMSSM  parameter $\mu_{\rm eff}$, further illustrating the correlation between model parameters and signal strengths.

\begin{figure}[!hptb]
    \centering
    \includegraphics[width=0.99\linewidth]{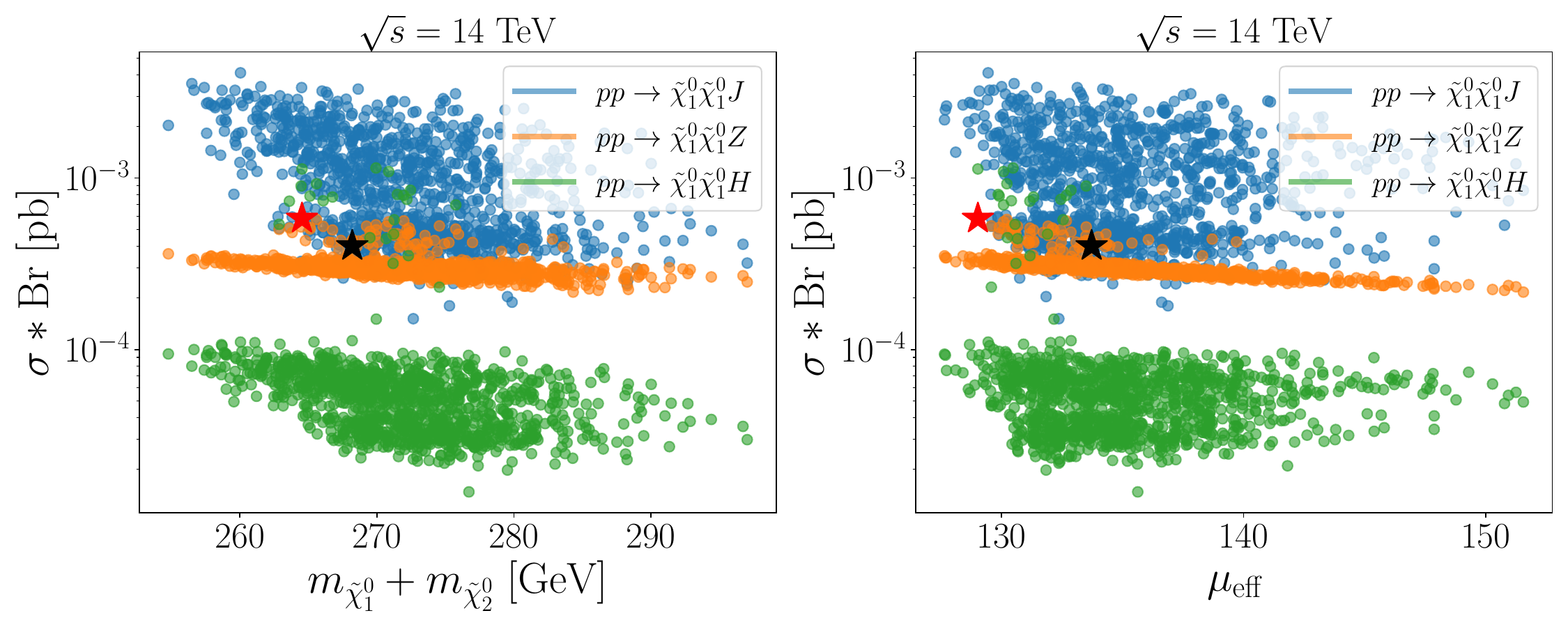}
    \caption{Left: Total production cross-sections for mono-jet (blue), mono-$Z$ (orange) and mono-$H$ (green) channels as a function of the combined mass of the LSP and NLSP neutralinos. Right: cross-sections versus the NMSSM parameter  $\mu_{\rm eff}$. In both plots, the red and black points correspond to the chosen BPs for the mono-$Z$ and -$H$ analyses, respectively.}
    \label{fig:sigma_mono}
\end{figure}

The full set of NMSSM input parameters for the selected BPs is provided in Table~\ref{table:nmssm-params}. These points are chosen to optimize the signal significance while satisfying existing experimental constraints from DM relic density, direct detection limits and collider bounds.

\begin{table}[!hptb]
\centering
\renewcommand{\arraystretch}{1.9}
\resizebox{\textwidth}{!}{%
\begin{tabular}{|c|c|c|c|c|c|c|c|c|c|c|c|}
\hline
 & $\tan\beta$ & $\lambda$ & $\kappa$ & $A_\lambda$ & $A_\kappa$ & $\mu_{\rm eff}$ & $M_1$ & $M_2$ & $M_3$ & $A_t$ &  $m_{\tilde{\chi}^0_1}$ \\
\hline
mono-$H$ & $4.62$ & $2.6\times 10^{-1}$ & $2.0\times 10^{-1}$ & $6.1\times 10^{2}$ & $-2.97\times 10^{2}$ & $1.33\times 10^{2}$ & $2.7\times 10^{3}$ & $1.08\times 10^{3}$ & $3\times 10^{3}$ & $-4.9\times 10^{3}$ & $1.26\times 10^{2}$ \\
\hline
mono-$Z$ & $4.62$ & $2.9\times 10^{-1}$ & $2.3\times 10^{-1}$ & $5.6\times 10^{2}$ & $2.94\times 10^{2}$ & $1.29\times 10^{2}$ & $2.49\times 10^{3}$ & $9.1\times 10^{2}$ & $1.4\times 10^{3}$ & $-4.6\times 10^{3}$ & $1.2\times 10^{2}$ \\
\hline
\end{tabular}
}
\caption{NMSSM input parameters for the BPs used in the mono-$H$ and -$Z$ analyses.}
\label{table:nmssm-params}
\end{table}
%Although we consider different analyses topologies but the background channels remain the same. 

\subsection{Background}

Having selected the BPs for the signal, we now turn to the analysis of the corresponding background processes. The most significant background arises from top quark production, including both \( t\bar{t} \) and single top processes. In particular, \( t\bar{t} \) production contributes to the final state with \( b\bar{b} \) pairs through its hadronic, semileptonic and leptonic decay modes. Among these, the semileptonic decay dominates due to its relatively high production rate and the presence of substantial $\slashed{E}_T$. The leptonic decay mode, while less prominent, also contributes non-negligibly, with \( \slashed{E}_T \) arising from the two neutrinos produced in leptonic \( W \) decays. However, the application of a lepton veto with a transverse momentum threshold of \( p_T > 10 \)~GeV effectively reduces this background at the selection level. In contrast, the semileptonic background is less suppressed by such a veto. The hadronic decay of \( t\bar{t} \), despite having the largest cross-section, contributes primarily through fake \( \slashed{E}_T \) originating from jet energy mismeasurement. A detailed simulation shows that the hadronic \( t\bar{t} \) background plays only a sub-dominant role. The single top background is also considered, though its impact remains minor due to a much lower production cross-section compared to the semileptonic and leptonic \( t\bar{t} \) channels.

Another significant source of background in the signal region is vector boson production in association with jets, denoted as \( V+\)jets (where \( V = W^\pm, Z \)). These processes feature large production cross-sections of order \( \sim 10^3~\mathrm{pb} \) and can generate considerable \( \slashed{E}_T \) via the semileptonic decays of the weak bosons. However, these backgrounds rely on the misidentification of two light-flavor jets as \( b \)-jets, a scenario with a low double-mistag probability of approximately 0.04\% as estimated from our Monte Carlo simulations. Notably, the contribution from \( W+\)jets is found to be sub-dominant relative to \( Z+\)jets. This is primarily due to the higher \( \slashed{E}_T \) typically observed in \( Z+\)jets events and the suppression of \( W^\pm+\)jets through the lepton veto.

The QCD production of $ b\bar{b} $ pairs presents a major challenge for this analysis, given its extremely large cross-section of around $10^5~\mathrm{pb}$. However, the impact of this background is highly sensitive to the $\slashed{E}_T$, which in this case arises from jet energy mismeasurement. By optimizing the cut on $\slashed{E}_T$ this background becomes less relevant relative to those from \( t\bar{t} \) and \( Z+\)jets. In this analysis we do not consider this background. 

Another irreducible background also comes from the process $ Z(\rightarrow \nu\bar{\nu})h_{\rm SM}(\rightarrow b\bar{b}) $. This process yields a final state with both significant $\slashed{E}_T$ and a $b\bar{b}$ invariant mass distribution similar to that of the signal. However, due to its comparatively small cross-section, approximately 100~fb, this background remains less significant than the others discussed above.
\begin{table}[!ht]
\centering
\renewcommand{\arraystretch}{1.2} % Optional: more vertical spacing
\begin{tabularx}{\textwidth}{|X|c|c|c|c|c|}
\hline
{Selection cuts} & $\bar{t}t$ leptonic & $V+jj$ & $VV$ & $Zh_{\rm SM}$ & mono-$H$ \\
\hline
Cross-section [pb] & 26.9 & 516 & 0.587 & 0.106 & $4\times 10^{-4}$ \\
\hline
$\slashed{E}_T\ge 200$~GeV & $1.5\times 10^{-2}$ & $1.5\times 10^{-3}$ & $2\times 10^{-2}$ & $4.7\times 10^{-2}$ & 0.436 \\
\hline
$N(b) \ge 2$ & $1.2\times 10^{-2}$ & $1.1\times 10^{-3}$ & $1.4\times 10^{-2}$ & $3.5\times 10^{-2}$ & 0.325 \\
\hline
$p_T(b_1)\ge 50 \ ,  \ p_T(b_2)\ge 30$~GeV & $1.8\times 10^{-3}$ & $1.7\times 10^{-4}$ & $3\times 10^{-3}$ & $1.4\times 10^{-2}$ & 0.13 \\
\hline
80~GeV $\le m(b\bar b) \le 140$~GeV & $5\times 10^{-4}$ & $6\times 10^{-5}$ & $2.6\times 10^{-3}$ & $1.2\times 10^{-2}$ & 0.121 \\
\hline
$\Delta\phi(b_{1,2},\slashed{E})> 0.35$ & $4\times 10^{-4}$ & $4\times 10^{-5}$ & $2.4\times 10^{-3}$ & $1.1\times 10^{-2}$ & 0.120 \\
\hline
\end{tabularx}
\caption{Signal and background efficiencies in the $H (\to b\bar{b})+ \slashed{E}_T$ final state after applying selection cuts at 14 TeV. Selection efficiencies are computed as the fraction of events remaining after applying the selection cut.}
\label{tab:cut_monoH}
\end{table}
\begin{table}[!ht]
\centering
\renewcommand{\arraystretch}{1.2} % Optional: more vertical spacing
\begin{tabularx}{\textwidth}{|X|c|c|c|c|c|}
\hline
{Selection cuts} & $\bar{t}t$ leptonic & $V+jj$ & $VV$ & $Zh_{\rm SM}$ & mono-$Z$ \\
\hline
Cross-section [pb] & 26.9 & 516 & 0.587 & 0.106 & $5\times 10^{-4}$ \\
\hline
$\slashed{E}_T\ge 200$~GeV & $1.5\times 10^{-2}$ & $1.5\times 10^{-3}$ & $2\times 10^{-2}$ & $4.7\times 10^{-2}$ & 0.38 \\
\hline
$N(b) \ge 2$ & $1.2\times 10^{-2}$ & $1.1\times 10^{-3}$ & $1.4\times 10^{-2}$ & $3.5\times 10^{-2}$ & 0.29 \\
\hline
$p_T(b_1)\ge 30 \ ,  \ p_T(b_2)\ge 20$~GeV & $1.1\times 10^{-2}$ & $5.5\times 10^{-4}$ & $4.8\times 10^{-2}$ & $2.8\times 10^{-2}$ & $0.182$ \\
\hline
50~GeV $\le m(b\bar b) \le 120$~GeV & $6\times 10^{-3}$ & $3\times 10^{-4}$ & $4.3\times 10^{-2}$ & $2.6\times 10^{-2}$ & $0.12$ \\
\hline
$\Delta\phi(b_{1,2},\slashed{E})> 0.35$ & $5.5\times 10^{-3}$ & $3\times 10^{-4}$ & $2.5\times 10^{-2}$ & $1.1\times 10^{-2}$ & $0.11$ \\
\hline
\end{tabularx}
\caption{Signal and background efficiencies in the $Z(\to b\bar{b})+ \slashed{E}_T$ final state after applying selection cuts at 14 TeV. Selection efficiencies are computed as the fraction of events remaining after applying the selection cut.}
\label{tab:cut_monoZ}
\end{table}

Based on the kinematics of signal and backgrounds, in order to generate the events, more effectively, we apply the following cuts at the simulation/generation (i.e., parton) level: $p_T ({\rm jet})\ge 20$~GeV, $p_T (l)\ge 20$~GeV, $p_T (b)\ge 20$, MET $\ge 30$~GeV and  $|\eta (l, {\rm jet}, b)| \le 2.5$ (hereafter, $j=e,\mu$).   Moreover,  we require the events to have at least two $b$-jets with cone radius $R=0.4$ using a flat tagging efficiency of $75\%$ where as for the mis-tagging efficiency of gluon- and light-quark-jets as $b$-ones, we adopt a flat rate of $10^{-3}$. After generating the events we reconstruct nine kinematic distributions for both signal and backgrounds as detailed below. To collect the two $b$-jets from the Higgs boson decay we do not require any mass window cut rather we collect the $b$-jet pair that has the closest invariant mass to the SM-like Higgs boson mass. 

Selection cuts and the corresponding signal and backgrounds efficiency for mono-$H$ and -$Z$ are reported in Tables \ref{tab:cut_monoH} and \ref{tab:cut_monoZ}, respectively. The selection cuts are optimized to maximize the signal-to-background yield resulting in different ranges for both analyses. 

%%================================+%%
%%================================+%%
\subsection{Analysis Setup and Data Structure}
%%================================+%%
%%================================+%%
To exploit the differences in QCD color flow between signal and background processes, we incorporate both reconstructed high-level kinematic variables and color flow information as input features to a multi-modal DL network. This approach allows the model to learn complementary aspects of the event topology: one related to momentum-space features and the other rooted in the QCD dynamics governing the QCD color flow.

The color flow of a QCD process is dictated by the underlying $SU(3)$ color symmetry, which enforces color conservation at each vertex of the interaction. The way color indices are contracted determines the radiation pattern observed in the final state. For example, in processes where jets originate from a color singlet, such as a Higgs or $Z$ boson, the color indices of the resulting quarks are contracted with each other, forming a color dipole. This typically results in soft hadronic radiation between the two jets. In contrast, for jets originating from the decay of a colored particle, such as a top quark, the color indices of the outgoing quarks are connected to those of the initial-state protons. This leads to isolated color poles and a soft radiation pattern that aligns more closely with the beam directions~\cite{Maltoni:2002mq,Hagiwara:2010vk,Kilian:2012pz}.

The global color flow structure manifests itself in the detector as a color string, a flux tube of soft hadrons  connecting color correlated final-state particles. In particular, the distribution of soft hadrons surrounding hard jets encodes valuable information about the color connections of the event~\cite{Kim:2019wns,Hammad:2022lzo}. This soft radiation pattern is sensitive to whether the two jets are color-connected to each other or to the beam remnants and can thus serve as a powerful discriminator between signal and background topologies.
In our case, the signal consists of two $b$-quarks arising from the hadronic decay of a color singlet  boson. These quarks form a color dipole and their associated radiation tends to fill the region between them. Conversely, in the dominant $t\bar{t}$ background, the two $b$-quarks are produced from the decay of color triplet top quarks. These $b$-quarks are individually color-connected to the incoming beams, resulting in two largely isolated radiation cones and a depletion of soft hadronic activity between them.
To capture this contrast, we design a multi-modal DL network with two distinct input streams: one capturing high-level kinematic variables and the other designed to characterize color flow features. For the latter, we utilize observables that are sensitive to soft radiation patterns surrounding the $b$-jets, such as energy flow and the transverse momentum density of all charged and neutral hadrons. Specifically, we consider six features for each final-state hadron: transverse momentum, energy, mass, charge, pseudorapidity and azimuthal angle.
These features are effective in capturing the differences in QCD color flow between signal and background events.

To complement the color flow observables, we use a set of high-level reconstructed kinematic variables as inputs to the DL network. These variables are chosen based on their ability to distinguish signal events from the dominant  backgrounds. The list of kinematic observables is as follows:

\begin{itemize}
    \item {MET or ${\not\hspace*{-0.1cm}E}_T$:} This  is defined as 
    \[
    {\rm MET} = - \left| \sum_{v_i} \vec{p}_T(v_i) \right|,
    \]
    where the sum runs over all visible particles. Signal events involving neutralino DM, with mass around 100~GeV, typically exhibit large MET due to the escaping LSPs, which helps to suppress $t\bar{t}$ and di-boson backgrounds.

    \item \textbf{$p_T(b_1)$:} Transverse momentum of the leading $b$-jet. In the signal process, the $b$-jets originate from the decay of a boosted SM boson resulting from a heavy mediator. This results in a $p_T$ spectrum that closely resembles that from $t\bar{t}$ events (where $b$-jets arise from top decays), but differs significantly from other background processes.

    \item \textbf{$p_T(b_2)$:} Transverse momentum of the subleading $b$-jet. 

    \item \textbf{$\eta(b_1)$:} Pseudorapidity of the leading $b$-jet. While this variable shows similar distributions for signal and most backgrounds, the $pp \to Zb\bar{b}$ process exhibits a broader $\eta$ distribution.

    \item \textbf{$\eta(b_2)$:} Pseudorapidity of the subleading $b$-jet, generally following the same trend as $\eta(b_1)$ across all processes.

    \item \textbf{$p_T(b\bar{b})$:} Transverse momentum of the $b$-jet pair that best reconstructs the  boson mass, either Higgs or $Z$. This variable highlights the boost of the Higgs or $Z$ boson in the signal events. 

    \item \textbf{$\eta(b\bar{b})$:} Pseudorapidity of the $b$-jet pair. 

    \item \textbf{$\Delta R(b,\bar{b})$:} Angular distance between the two $b$-jets reconstructing the Higgs or $Z$ boson, defined as
    \[
    \Delta R(b,\bar{b}) = \sqrt{(\Delta\eta(b,\bar{b}))^2 + (\Delta\phi(b,\bar{b}))^2}.
    \]
    In $ZZ$ and $hZ$ background processes, where the $b$-jets come from the decay of a gauge or Higgs boson produced near resonance, the jets are mostly back-to-back, peaking at $\Delta R (b,\bar b)\sim 3$.  For QCD-induced $Zb\bar{b}$ events, the angular separation has a broader distribution. In contrast, signal events with boosted  bosons produce $b$-jets with smaller separation.

    \item \textbf{$m(b\bar{b})$:} Invariant mass of the $b$-jet pair closest to the Higgs or $Z$ boson mass. 
    \end{itemize}

To prepare the data for DL training, we construct two distinct datasets.
The first dataset captures QCD color flow information and has the shape $(N, 80, 6)$, where $N$ is the number of training events, 80 is the fixed number of hadrons considered per event and 6 is the number of features for each hadron.  For events with more than 80 hadrons, only the 80 with the highest momenta are selected; for events with fewer hadrons, the remaining slots are zero-padded.
The second dataset encodes high-level kinematic variables of the final-state particles and has the shape $(N, 9)$, where the 9 input features are as described earlier.
To ensure balanced training, we use equal-sized datasets for signal and background events, with $N = 600{,}000$ for each class. Since this is a supervised classification task, we assign binary labels: $Y=1$ for signal events and $Y=0$ for background events.
During training, the network iteratively adjusts its weights to minimize classification error, learning to correctly associate the input features with their corresponding labels until satisfactory accuracy is achieved.

For event simulation, we  employ \texttt{MadGraph5}~\cite{Alwall:2014hca,Frederix:2018nkq} for cross-section estimation and generating parton-level events. Then  \texttt{Pythia8.3}~\cite{Bierlich:2022pfr} is utilized to include parton showering and hadronization effects. The factorization and renormalization scales have been kept at the default {\tt MadGraph} event by event dynamic choice. Jets are formed using \texttt{FastJet} package~\cite{Cacciari:2011ma} utilizing anti-KT algorithm~\cite{Cacciari:2008gp} with $R=0.4$. Detector effects are taken into account with the \texttt{Delphes} package~\cite{deFavereau:2013fsa} using the default ATLAS card.
For the DL analysis, we use \texttt{PyTorch Geometric}~\cite{fey2019fast} for building the GNN layers, while standard \texttt{PyTorch}~\cite{paszke2019pytorch} is used for the MLP. Finally, the  \texttt{Scikit-Learn} package~\cite{pedregosa2011scikit} is used to facilitate network training and evaluation. 

%%================================+%%
%%================================+%%
\subsection{DL Network Structure}
%%================================+%%
%%================================+%%
To effectively analyze two distinct input datasets, we employ a multi-modal NN that combines a Graph Convolutional Network (GCN) with a Multi-Layer Perceptron (MLP). The GCN component is particularly well-suited for learning from graph-structured data, as it incorporates topological relationships among nodes and edges, enabling the model to capture spatial and relational patterns.
In our framework, we represent the final-state hadrons as a fully connected graph, where each node corresponds to a hadron and is described by a feature vector $x = (M, p_T, E, \eta, \phi, \rm{charge})$. This vector encapsulates the properties of the corresponding particle, where $p_T$ denotes transverse momentum, $E$ is the energy, $\eta$ represents pseudorapidity and $\phi$ is the azimuthal angle. The graph is fully connected, with edges weighted by the angular distance $\Delta R_{(x_{i}, x_{j})}$ between the particles in nodes $i$ and $j$, thereby enabling the model to capture intricate spatial relationships between such particles.
The primary objective of the GCN is to learn a function that maps the input features to new representations, effectively capturing the complex relationships between the graph nodes~\cite{Bachlechner:2022cvf,Esmail:2023axd,Sahu:2024sts}. The core idea behind GCNs is the extension of the convolution operation from regular grid structures, as in CNNs, to irregular graph domains. A graph convolution can be interpreted as a local aggregation of information from neighboring nodes, combining both the structural properties of the graph and the features associated with each node. Given an input graph $G = (V, E)$, the graph convolution operation is defined as
\begin{equation*}
H^{(l+1)} = \sigma \left( \hat{D}^{-\frac{1}{2}} \hat{A} \hat{D}^{-\frac{1}{2}} H^{(l)} W^{(l)} \right),
\end{equation*}
where  $H^{(l)} \in \mathbb{R}^{N \times F_l} $ denotes the feature matrix at layer $l$, with  $N$ being the number of vertices in the graph and $F_l$ the dimensionality of the feature space at that layer. The matrix  $W^{(l)} \in \mathbb{R}^{F_l \times F_{l+1}}$ is the trainable weight matrix at layer $l$ and $\sigma$ represents the non-linear activation function.
The adjacency matrix of the graph, is given by $\hat{A} \in \mathbb{R}^{N \times N}$, defined as $\hat{A} = A + I_N $, where  $A$ is the original adjacency matrix and $ I_N$ is the identity matrix of size $N$. The diagonal degree matrix corresponding to $\hat{A}$ is denoted by $ \hat{D} \in \mathbb{R}^{N \times N} $, with elements $\hat{D}_{ii} = \sum_j \hat{A}_{ij}$, which represent the number of connected edges to vertex $i $.
The graph convolution operation can be interpreted as a message-passing scheme, where each node aggregates information from its neighbors and updates its feature representation using the learned weights. By stacking multiple GCN layers, the model is able to learn higher-order interactions among nodes, thereby capturing increasingly complex relational structures within the graph representing the color flow for different processes. 
The second stream of the network, which processes the kinematic information, begins with an input layer of nine neurons. This is followed by a series of fully connected layers. The outputs from both streams are then concatenated and passed to the final output layer for classification. Illustration of the schematic architecture of the network is shown in Figure \ref{fig:network}.
\begin{figure}[!tphb]
    \includegraphics[width=0.99\linewidth]{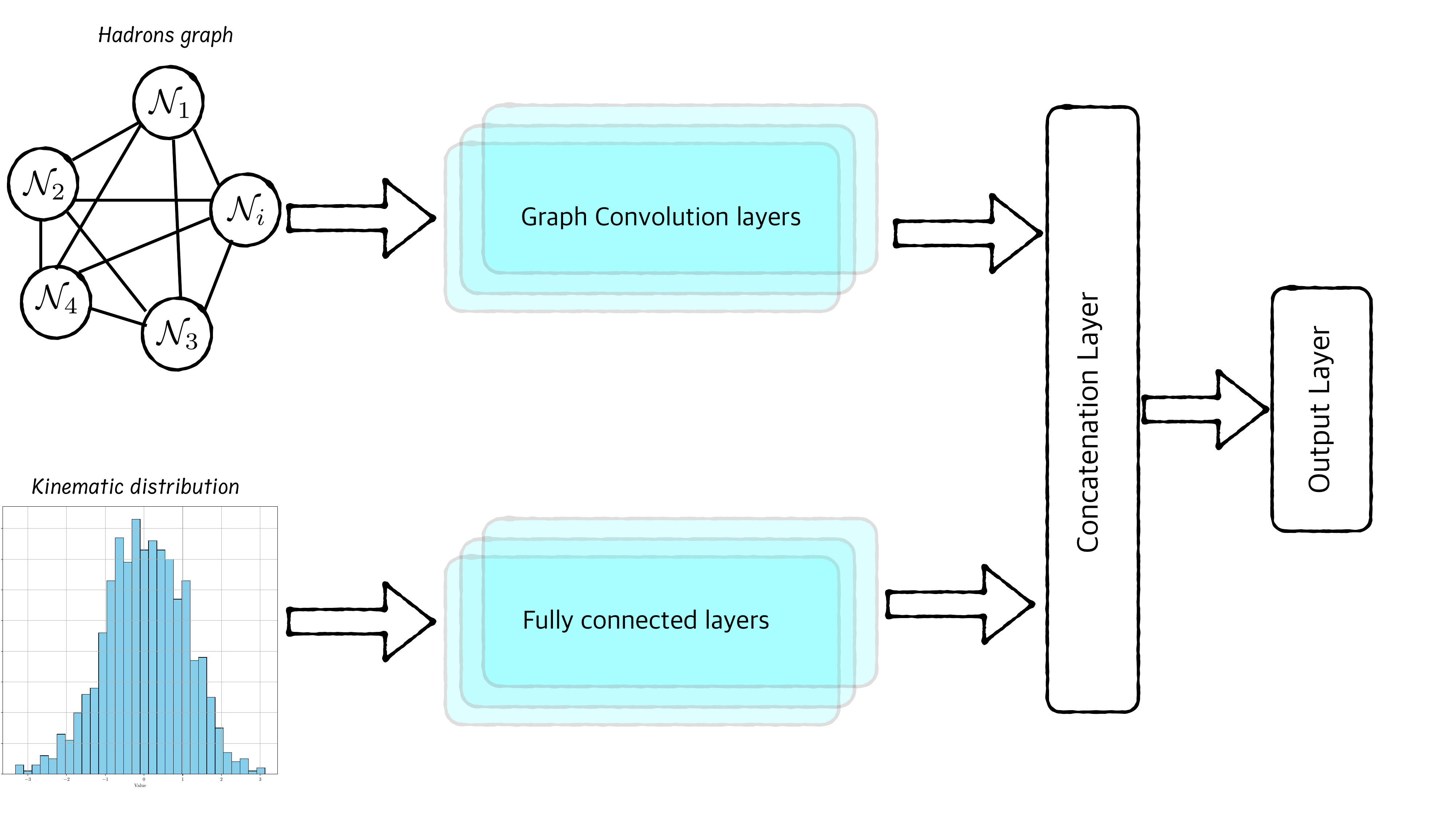}
    \caption{Schematic diagram of the multi-modal GNN architecture, illustrating the data flow from the input graph of hadrons through the GCN to produce intermediate representations. These representations are then concatenated with high-level kinematic variables and passed through an MLP, resulting in the final classification output.}
    \label{fig:network}
\end{figure}
While a sufficiently deep GNN could, in principle, capture global event properties from low level hadronic inputs, kinematic variables generally provide stronger discriminating power in BSM searches. Combining both types of information leads to a clear improvement in classification performance, as demonstrated in Ref.~\cite{Hammad:2023sbd}, where the joint use of hadronic and kinematic information outperformed models trained on either input alone.

Designing a DL network involves specifying its architecture, which includes a number of parameters whose values must be determined before training. These are known as hyperparameters and, unlike model weights, they cannot be learned during training and must be set manually. The performance of the network is highly sensitive to the choice of these hyperparameters, making their careful selection essential for achieving optimal results. A common method for hyperparameter tuning is grid search, one of the most straightforward techniques. In this approach, the model is trained and evaluated for every possible combination of predefined hyperparameter values. The combination that yields the best performance is then selected. While grid search can be effective in identifying optimal configurations, it is computationally expensive and often impractical for large search spaces. To address this limitation, we adopt a more efficient alternative known as random search. Instead of exhaustively testing all combinations, random search samples hyperparameter values from specified probability distributions. This significantly reduces the computational cost while still allowing for the discovery of well-performing configurations, especially when only a subset of hyperparameters strongly influences performance. After hyperparameters optimization the best network structure consists of 3 GCN layers with ReLU activation function. This is followed by max pooling to aggregate the node embedding. The other part of the model is a basic set of 3 fully connected layers with number of neurons of 128, 512 and 256. Output from both streams are concatenated in on layer and passed to an output layer with two neurons and softmax function. 
%%================================+%%
%%================================+%%
\subsection{Sensitivity at the (HL-)LHC}
%%================================+%%
%%================================+%%
After preparing the datasets, we train the networks to learn the complex, non-linear mapping between the input features and their corresponding labels. Signal events are assigned the label $Y = 1$, while background events are labeled $Y = 0$. To eliminate any bias related to the ordering of signal and background events, we combine both classes into a single dataset and shuffle the events along with their assigned labels. During training, the network proceeds through multiple epochs, where each epoch is defined as one complete pass over the entire dataset. For every iteration, the network updates its weights using backpropagation, which propagates the prediction error backward to adjust the network parameters. The training objective is to minimize a loss function that quantifies the discrepancy between predicted outputs and the true labels, thereby guiding the network toward an optimal set of weights. In this network we  employ the categorical cross-entropy loss function, defined as $ -\sum_i Y_i \log (\hat{Y}_i)$, where $i = 0, 1$ corresponds to the background and signal classes, respectively, plus $Y_i $ and $ \hat{Y}_i$ denote the true and predicted label probabilities, respectively.  Training is conducted over 22 epochs with a batch size of 500 samples. The final network output is a probability vector $\hat{Y} $ of dimension $1 \times 2$, given by $(\mathcal{P}_{\text{sig}}, \mathcal{P}_{\text{bkg}}) $, where $\mathcal{P} \in [0,1] $. An event is classified as signal if $ \mathcal{P}_{\text{sig}} > 0.5 $ (equivalently, \( \mathcal{P}_{\text{bkg}} < 0.5 \)) and as background if \( \mathcal{P}_{\text{sig}} < 0.5 \) (i.e., \( \mathcal{P}_{\text{bkg}} > 0.5 \)).
Once training is complete, the model performance is evaluated on a completely unseen test dataset to assess its generalization capability.
\begin{figure}[!tphb]
\centering
    \includegraphics[width=0.99\linewidth]{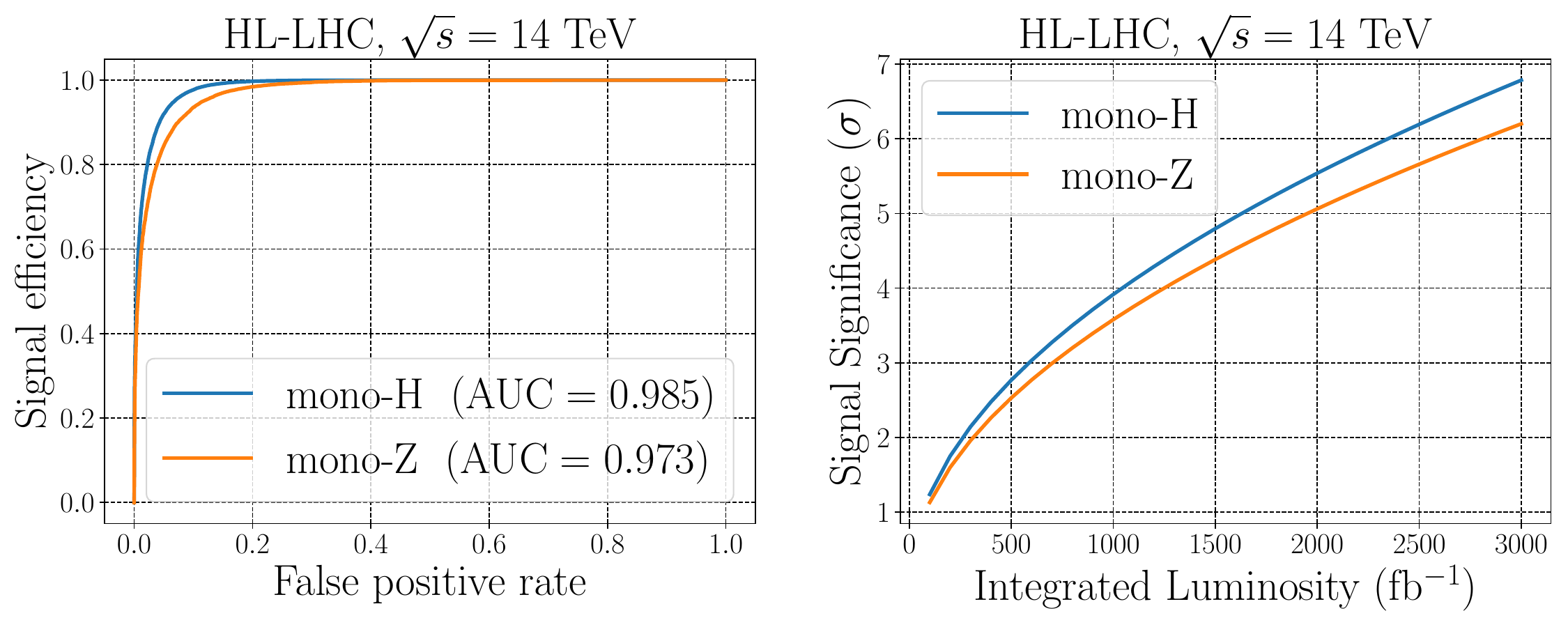}
    \caption{Left: ROC curves for the mono-$H$ analysis in blue and the mono-$Z$ analysis in orange. Right: signal significance at the (HL-)LHC for an integrated luminosity ranging from 100 to 3000~fb$^{-1}$. }
    \label{fig:ROC}
\end{figure}

The effectiveness of each network in distinguishing signal from background events depends on how well it captures and disentangles the underlying physical features encoded in complex kinematic distributions and color flow patterns. This can be quantified using different metrics such as the Receiver Operating Characteristic (ROC) curves and  the Area Under the Curve (AUC), which measures the model ability to correctly differentiate between signal and background events across varying decision thresholds. Figure \ref{fig:ROC}, left plot, shows the ROC curves for the Mono-$H$ and -$Z$ analyses. The mono-$H$ analysis has a better performance, with AUC $=0.985$, over the mono-$Z$ analysis, which has AUC $=0.973$. The signal significances as a function of the network output are computed using the formula~\cite{LHCDarkMatterWorkingGroup:2018ufk,Antusch:2018bgr}:
\begin{equation}
\sigma = \left[ 2\left( (N_s+N_b)\ln\frac{(N_s+N_b)(N_b+\sigma^2_b)}{N_b^2+(N_s+N_b)\sigma^2_b}  -\frac{N^2_b}{\sigma^2_b}\ln(1+\frac{\sigma^2_b N_s}{N_b(N_b+\sigma^2_b)})         \right) \right]^{1/2}\,,
\end{equation}
with $N_s$, $N_b$ being the number of signal and background events, respectively, and $\sigma_b$ parameterizing the systematic  uncertainty on the latter which is considered as $10\%$. The right plot of Figure~\ref{fig:ROC} displays the expected signal significance as a function of the integrated luminosity. At a projected luminosity of $\mathcal{L} = 3000~\mathrm{fb}^{-1}$, the mono-$Z$ channel reaches a significance of $6.1\sigma$, while the mono-$H$ channel attains $6.8\sigma$. These results  highlight the potential of probing joint anomalies linking the existing LHC excesses, as discussed in Section  3, with the mono-$H$ and -$Z$ channels, by the end of the HL-LHC run.
%%%%%%%%%%%%%%%%%%%%%%%%%%%%%%%%%%%%%%%%%%%%%%%%%%%%%%%%%%%%%%%
\section{Results and Discussion}%%%%%%%%%%%%%%%%%%%%%%%%%%%%%%
\label{sec:results}
%%%%%%%%%%%%%%%%%%%%%%%%%%%%%%%%%%%%%%%%%%%%%%%%%%%%%%%%%%%%%%%
Through an extensive scan of the NMSSM parameter space, aimed at accommodating all current theoretical and experimental constraints as well as various observed anomalies, we have identified a ``golden region'' of it that satisfies all such requirements. Specifically, leveraging this result, we have demonstrated that this region can naturally account for the observed 95~GeV and 650~GeV excesses while also providing a plausible interpretation of the reported EWino excess, 
including being consistent with the most recent measurement of $(g-2)_\mu$. Over such a portion of the scanned parameter space, we have found agreement between theoretical predictions and experimental data at the $2\sigma$ level.  

This result has been aided by the use of a dedicated DL-informer parameter scanner.
Initially, the DL scan explored broad parameter ranges, which were subsequently narrowed to accommodate the combined constraints and anomalies. The final accepted points constitute approximately 10\% of the collected points from the narrow-range scan. Specifically, 1179 points were collected from this narrower scan, all of which satisfy all bounds and anomalies except for the 95~GeV and 650~GeV excesses. Imposing further the requirements for the  95~GeV and 650~GeV anomalies  eliminates 90.33\% of the collected points. The ranges of the twelve scanned parameters for the remaining allowed points are summarized in Table~\ref{tab:allowed}.

\begin{table}[!ht]
\begin{center}
\begin{tabular}{ccccc}
 \toprule
  & $\tan\beta$ & $\lambda$ & $\kappa$ & $A_\lambda$ \\
  &  [4.33 , 4.92] & [0.22 , 0.31] & [0.16 , 0.23] & [538.11 , 624.63] \\
    \midrule
  & $A_\kappa$ & $\mu_\text{eff}$ & $M_1$ & $M_2$ \\
   & [$-299.58$ , $-220.86$] & [129.03 , 139.21] & [557.98 , 2984.42] & [915.37 , 4459.53] \\
    \midrule
   & $M_3$ & $A_t$ & $M_{Q_3}$ & $M_{U_3}$ \\
    &[1037.52 , 4696.11] & [$-4985.21$ , $-4129.13$] & [939.69 , 4333.26] & [1150.53 , 4756.70] \\
 \bottomrule
\end{tabular}
\end{center}
\caption{\label{tab:allowed}
    Allowed ranges of the scanned parameters that satisfy all the stated anomalies and constraints within the $2\sigma$ level.}
\end{table}
 To further quantify our results, we have evaluated the $\chi^2$ values for the final set of accepted points, incorporating all the constraints and anomalies discussed above. The combined $\chi^2$ is calculated using the standard expression: $\chi^2 = \sum_i \frac{\left(\rm{observed}_i - \rm{expected}_i \right)^2}{\sigma_i^2}$,
 %\begin{eqnarray}
  %  \chi^2 = \sum_i \frac{\left(\rm{observed}_i - \rm{expected}_i \right)^2}{\sigma_i^2}\,,
%\end{eqnarray}
where $\sigma_i$ denotes the uncertainty associated with the observed measurement.  

\begin{figure}[!h]
    \includegraphics[width=0.72\paperwidth,height=0.67\paperheight]{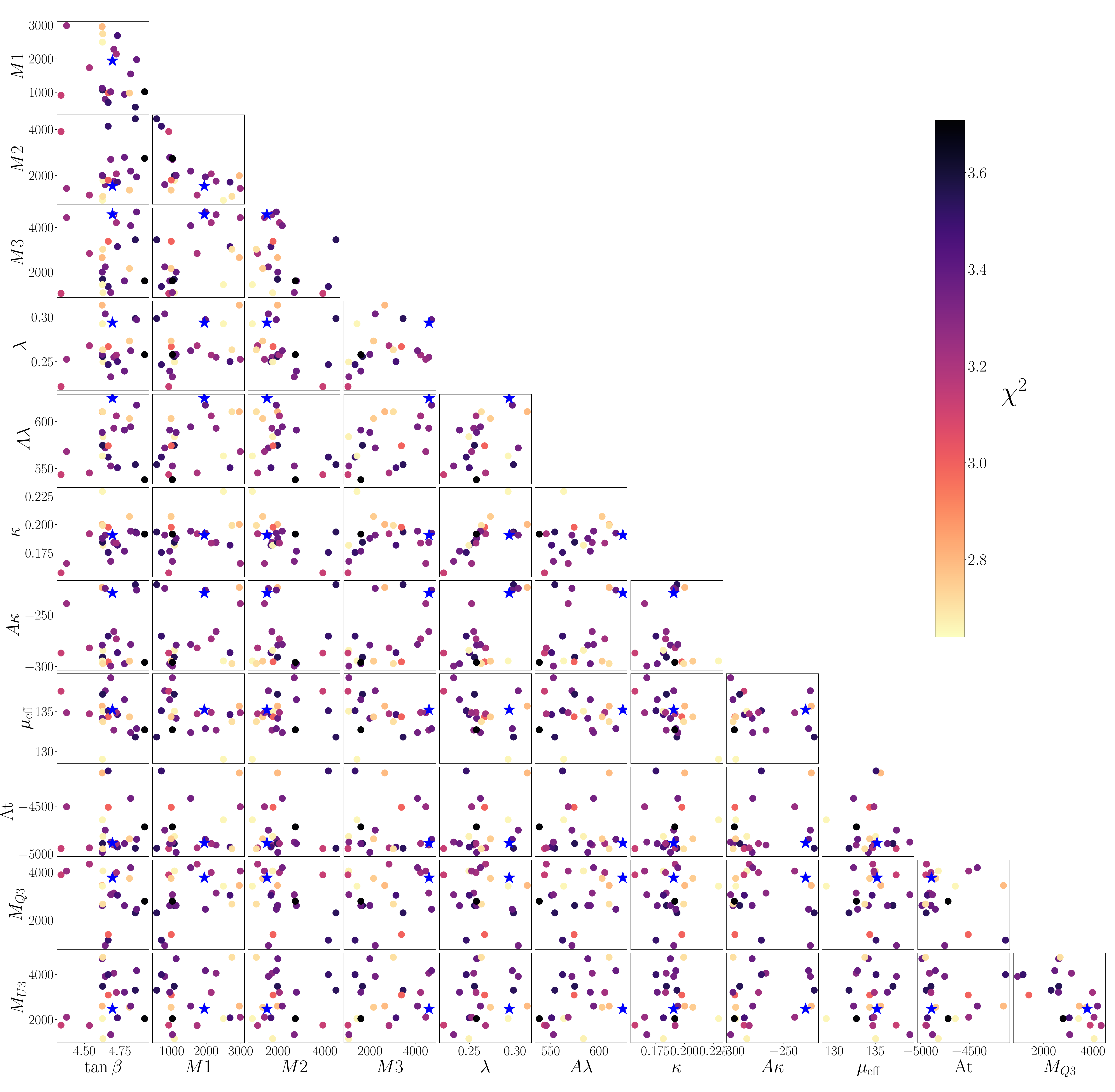}
   \caption{Ranges of the scanned variables in this analysis, with the color bar indicating the combined $\chi^2$ values. The minimum $\chi^2$ value is $2.63$, marked by a blue star.}
   
    \label{fig:chisquared}
\end{figure}

\begin{table}[!h]
\begin{center}
\begin{tabular}{ccccccc}
 \toprule
  & $\tan\beta$ & $\lambda$ & $\kappa$ & $A_\lambda$ & $A_\kappa$ & $\mu_\text{eff}$\\
   &  4.697 & 0.294 & 0.191 & 624.626 & $-228.919$ & 135.211 \\
    \midrule
   &  $M_1$ & $M_2$  & $M_3$ & $A_t$ & $M_{Q_3}$ & $M_{U_3}$ \\
   & 1940.590 &1535.770 &4584.030 & $-4885.850$ & 3764.090 & 2473.020 \\
 \bottomrule
\end{tabular}
\end{center}
\caption{\label{tab:bestfit}
    Values of the scanned parameters corresponding to the best-fit point, which yields a minimum $\chi^2$ value of $2.63$. This point is illustrated by a blue star in Figure~\ref{fig:chisquared}.  }
\end{table}

Figure~\ref{fig:chisquared} illustrates the distribution of the final accepted points across the scanned parameter space, with the color bar indicating the corresponding $\chi^2$ values. The blue star marks the point with the minimum $\chi^2$, for which detailed information is provided in Table~\ref{tab:bestfit}. According to our findings, this point is crucial and serves as a golden benchmark for experimental groups to pursue.

%%%%%%%%%%%%%%%%%%%%%%%%%%%%%%%%%%%%%%%%%%%%%%%%%%%%%%%%%%%%%%%
\section{Conclusions}%%%%%%%%%%%%%%%%%%%%%%%%%%%%%%%%%%%%%%%%%%%
%%%%%%%%%%%%%%%%%%%%%%%%%%%%%%%%%%%%%%%%%%%%%%%%%%%%%%%%%%%%%%%
In this paper, we have tested the possibility that an attractive realization of SUSY, the so-called NMSSM, can be the theoretical scenario underpinning some (current) experimental data that have revealed themselves anomalous to a reasonable degree of confidence. This choice has clear theoretical motivations. On the one hand, being a SUSY construct, the NMSSM can resolve the hierarchy problem of the SM while simultaneously generating the Higgs potential dynamically, predicting a low mass Higgs state, enabling coupling unification as well as providing a viable DM candidate. On the other hand, the NMSSM does not suffer from the `little hierarchy' and $\mu$ problems of the minimal SUSY realization, the MSSM, while simultaneously embedding enlarged Higgs and sparticle sectors, including a different DM candidate.

Armed with these features, we have exploited the NMSSM parameter space to interpret several excesses reported at the LHC, namely the 95 GeV and 650 GeV anomalies, as well as the so-called EWino excess observed by ATLAS and CMS. To this end, we have carried out an extensive scan of the NMSSM parameter space under the requirement that all theoretical and experimental constraints be satisfied, including the most recent measurement of the anomalous magnetic moment of the muon, $(g-2)_\mu$. This analysis has enabled us to identify a ``golden region'' of parameter space that simultaneously accommodates all such requirements.

A crucial ingredient in achieving these results has been the use of DL assisted scans. The NMSSM parameter space is intrinsically high-dimensional and tightly constrained by a set of correlated theoretical and experimental requirements, which makes exhaustive scans computationally prohibitive. Traditional random or grid-based sampling methods are typically inefficient in capturing finely tuned regions where several constraints overlap. In contrast, DL assisted approaches can efficiently learn the structure of the viable parameter space, guiding the exploration toward regions of phenomenological relevance while avoiding redundant sampling of disallowed configurations. This strategy has proven vital in uncovering the aforementioned golden region where multiple anomalies and constraints converge. In this sense, DL not only accelerates the scanning procedure but also improves its precision, enabling us to probe regions that would otherwise remain inaccessible to conventional techniques.

Through this approach, we have established agreement between theoretical predictions and experimental observations at the $2\sigma$ level across sizeable portions of the NMSSM parameter space. Building on this result, we predict that, within these regions, mono-$H$ (with $H = h_{\rm SM}$) and mono-$Z$ processes constitute promising channels for testing our hypothesis at the HL-LHC. Their distinctive kinematic features would provide a unique hallmark of such a non-minimal SUSY framework, offering a concrete path toward experimental validation in future collider searches.

Taken together, our results highlight the NMSSM as a theoretically robust and phenomenologically viable framework capable of addressing both longstanding theoretical challenges of the SM and a range of current experimental anomalies, while also demonstrating the power of modern DL assisted methodologies in tackling the complexity of high-dimensional SUSY parameter spaces.

Finally, 
to ease reproducibility, the code for the scanning implementation used in this analysis has been added to the \href{https://github.com/raalraan/DLScanner}{DLScanner GitHub repository} and is openly accessible in the
\href{https://github.com/raalraan/DLScanner/tree/main/tests/examples}{`examples'} directory under `tests'.

%%%%%%%%%%%%%%%%%%%%%%%%%%%%%%%%%%%%%%%%%%%%%%%%%%%%%%%%%%%%%%%
\section*{Acknowledgments}%%%%%%%%%%%%%%%%%%%%%%%%%%%%%%%%%%%%
%%%%%%%%%%%%%%%%%%%%%%%%%%%%%%%%%%%%%%%%%%%%%%%%%%%%%%%%%%%%%%%
SM is supported in part through the NExT Institute and the STFC Consolidated Grant ST/X000583/1. AH is funded by grant number 22H05113, ``Foundation of Machine Learning Physics'', Grant in Aid for Transformative Research Areas and 22K03626, Grant-in-Aid for Scientific Research (C). AH  is  partially supported by the Science, Technology and  Innovation Funding Authority (STDF) under grant number 50806.
The work of RR is supported by a KIAS Individual Grant (QP094601)
via the Quantum Universe Center at the Korea Institute for Advanced Study (KIAS). PK is supported by KIAS Individual Grant No. PG021403. 
Finally, this work used computational resources supported by the Center for Advanced Computation at KIAS. 
%%%%%%%%%%%%%%%%%%%%%%%%%%%%%%%%%%%%%%%%%%%%%%%%%%%%%%%%%%%%%%%
\section*{Dedication}%%%%%%%%%%%%%%%%%%%%%%%%%%%%%%%%%%%%
We would like to dedicate this paper to the late Prof.~Ulrich Ellwanger, one of the most passionate advocates of the NMSSM and a truly inspiring figure for us all. 
%%%%%%%%%%%%%%%%%%%%%%%%%%%%%%%%%%%%%%%%%%%%%%%%%%%%%%%%%%%%%%%
%%-------------------------%%
\bibliographystyle{JHEP}
\bibliography{biblo}
%%-------------------------%%
\end{document}